\documentclass[%
  reprint,
  prx,
  superscriptaddress,
  longbibliography,
  nofootinbib,
  amsmath,amssymb,
  aps,
]{revtex4-2}
\usepackage{color}
\usepackage{amsmath,amsthm,mathrsfs,dsfont}
\usepackage{graphicx}
\usepackage{tikz}
\usepackage{xr}
\usepackage{xc}
\usepackage{float}
\usepackage{subfigure}
\usepackage{bbm}
\usepackage{epsfig}
\usepackage{verbatim}
\usepackage{array}
\usepackage{setspace}
\usepackage{bbding}
\usepackage{amssymb}% http://ctan.org/pkg/amssymb
\usepackage{pifont}% http://ctan.org/pkg/pifont
\usepackage{braket}
\usepackage{ulem}
\usepackage{newtxtext,newtxmath}
\usepackage{enumerate}
\usepackage{thmtools}
\usepackage{ulem}

\newcolumntype{M}[1]{>{\centering\arraybackslash}m{#1}}
\newcolumntype{N}{@{}m{0pt}@{}}

\usepackage[unicode=true,
 bookmarks=true,bookmarksnumbered=false,bookmarksopen=false,
 breaklinks=false,pdfborder={0 0 1},backref=false,colorlinks=true]
 {hyperref}
\hypersetup{
 linkcolor=magenta, urlcolor=blue, citecolor=blue, pdfstartview={FitH}, hyperfootnotes=false, unicode=true}

\usepackage{color}
\makeatletter
\@ifundefined{textcolor}{}
{%
 \definecolor{BLACK}{gray}{0}
 \definecolor{WHITE}{gray}{1}
 \definecolor{RED}{rgb}{1,0,0}
 \definecolor{GREEN}{rgb}{0,1,0}
 \definecolor{BLUE}{rgb}{0,0,1}
 \definecolor{CYAN}{cmyk}{1,0,0,0}
 \definecolor{MAGENTA}{cmyk}{0,1,0,0}
 \definecolor{YELLOW}{cmyk}{0,0,1,0}
}

\usepackage{amsfonts}\usepackage{tabularx}\usepackage{dcolumn}\usepackage{bm}\usepackage{graphicx}\usepackage{epstopdf}

\DeclareMathOperator{\tr}{tr}

\hypersetup{urlcolor=blue}

\makeatother

\newcolumntype{C}[1]{>{\centering\arraybackslash$}p{#1}<{$}}

\DeclareMathOperator*{\minimize}{minimize}

\begin{document}

\widetext

%\title{Efficient and scalable suppression of heating errors for trapped ion quantum computing}
\title{Scalable suppression of heating errors in large trapped-ion quantum processors}
\author{Zixuan Huo}
\affiliation{Center on Frontiers of Computing Studies, School of Computer Science, Peking University, Beijing 100871, China}

\author{Yangchao Shen}
\affiliation{Center on Frontiers of Computing Studies, School of Computer Science, Peking University, Beijing 100871, China}

\author{Xiao Yuan}
% \email{xiaoyuan@pku.edu.cn}
\affiliation{Center on Frontiers of Computing Studies, School of Computer Science, Peking University, Beijing 100871, China}

\author{Xiao-Ming Zhang}
\email{xmzhang93@pku.edu.cn}
\affiliation{Center on Frontiers of Computing Studies, School of Computer Science, Peking University, Beijing 100871, China}
\affiliation {Key Laboratory of Atomic and Subatomic Structure and Quantum Control (Ministry of Education), Guangdong Basic Research Center of Excellence for Structure and Fundamental Interactions of Matter, School of Physics, South China Normal University, Guangzhou 510006, China}
\affiliation {Guangdong Provincial Key Laboratory of Quantum Engineering and Quantum Materials, Guangdong-Hong Kong Joint Laboratory of Quantum Matter, Frontier Research Institute for Physics, South China Normal University, Guangzhou 510006, China}

\begin{abstract}
Trapped-ion processors are leading candidates for scalable quantum computation. However, motional heating remains a key obstacle to fault-tolerant operation, especially when system size increases. Heating error is particularly challenging to suppress due to is incoherence nature, and no general methods currently exist for mitigating their impact in large systems with multiple phonon modes. In this work, based on a careful analysis about the dependence of heating-induced infidelity on phase-space trajectories, we present a simple yet comprehensive framework for suppressing heating errors in large trapped-ion quantum processors. Our approach is flexible, allowing various control pulse bases, ion numbers, and noise levels. Our approach is also compatible with existing error-mitigation techniques, including those targeting laser phase and frequency noise. Crucially, it relies on an efficiently computable cost function that avoids the exponential overhead of full fidelity estimation. We perform numerical simulations for systems with up to 55 qubits,
% demonstrating up to an order-of-magnitude reduction in infidelities.
demonstrating up to a fivefold reduction in infidelities compared with the conventional method in typical regimes.
We further show that our framework simultaneously reduces the sensitivity of the gate rotation angle to detuning errors, leading to an improvement in detuning-error tolerance. These results offer a practical route toward robust, large-scale quantum computation with trapped ions.
\end{abstract}

\maketitle

\section{introduction}
\label{sec:intro}
Quantum computing holds the potential to address problems that are intractable for classical computers. Among the various quantum platforms currently under development, trapped ion systems have emerged as a leading candidate in the pursuit of scalable and universal quantum machines~\cite{haffner2008quantum, monroe2013scaling,bruzewicz2019trapped}. This progress has been supported by a series of key experimental achievements, including single-qubit gate infidelities as low as $10^{-6}$~\cite{harty2014high} and two-qubit gate infidelities of $1.84\times10^{-3}$~\cite{moses2023race}, demonstrating the high level of precision now attainable with this technology. Moreover, the successful implementation of systems with up to 56 interconnected qubits~\cite{decross2024computational,moses2023race} reflects substantial advances in scaling capabilities. Beyond their performance in digital quantum computation, trapped ion systems have also established themselves as a versatile platform for analog quantum simulations, providing valuable opportunities for investigating complex many-body quantum phenomena~\cite{guo2024site, feng2023continuous,iqbal2024non,chertkov2023characterizing}.

Despite these advances, the realization of universal trapped ion quantum computation remains hindered by unavoidable noise arising from interactions with the environment. Among the various sources of error, heating noise represents a particularly significant challenge~\cite{ballance2016high, baldwin2021high}. Such noise originates from environmental perturbations of the ions’ motional states. As the system size increases, the number of motional modes—and consequently, the number of dissipation channels—also grows, amplifying the impact of heating errors~\cite{wu2018noise,he2024scaling}. In contrast to coherent errors, which can often be systematically corrected, the decoherent nature of heating noise makes it resistant to complete suppression.
Initial efforts to mitigate heating noise have shown that carefully engineered polychromatic driving can effectively reduce its influence in small systems~\cite{Haddadfarshi_2016, shapira2018robust, webb2018resilient}. However, these methods typically require complex experimental configurations and are restricted to models with only a single phonon mode, posing significant limitations to their scalability.

In this work, we present a flexible and practical framework for suppressing heating errors in trapped ion quantum computers. The proposed approach is scalable to any number of ions and supports Mølmer–Sørensen gates~\cite{molmer1999multiparticle,sorensen1999quantum} with arbitrary rotation angles. By optimizing control pulses to minimize an analytically derived objective function that quantifies heating-induced infidelity, this framework not only mitigates heating noise but also offers the potential to address additional error sources, such as mode frequency drift.
Taking standard amplitude modulation protocols as illustrative examples, we reformulate the pulse optimization problem as a positive-semidefinite quadratically constrained quadratic programming problem using the positive-extraction approximation, allowing for efficient numerical solution. Simulation results demonstrate that this method can reduce infidelity by
% an order of magnitude
up to fivefold in typical regimes
compared to conventional pulse designs, while maintaining practical pulse power requirements. Beyond heating-error suppression, our optimization framework simultaneously reduces the sensitivity of the gate rotation angle to detuning errors—an equally important aspect that has received limited attention in prior work.
As another interesting observation, some of our heating-error-optimized waveforms resemble the combination of alternating sign segments~\cite{hayes2012coherent, loschnauer2025scalable} and slowly varying envelope protocols~\cite{tyler2024laser, hughes2025trapped} designed for suppressing detuning-related errors. This result further suggest that the minimization of different error sources can be relevant in practice.

%Interestingly, the optimized waveform produced by our numerical procedure exhibits two qualitative features that resemble strategies previously used to suppress detuning-related errors: a slowly varying envelope and alternating sign segments. Similar structures have been discussed in earlier works~\cite{hayes2012coherent,tyler2024laser} in the context of reducing residual motional displacement.

The remainder of this paper is structured as follows. Section~\ref{sec:preli} reviews the fundamentals of Mølmer–Sørensen gates and outlines the criteria for frequency robustness. In Section~\ref{sec:suppression}, we present our framework for suppressing heating errors, including an efficient method for estimating heating-induced infidelity, the positive-extraction approximation, and the final optimization formulation. Section~\ref{sec:results} provides numerical results demonstrating the effectiveness of our approach. We conclude in Section~\ref{sec:conclusion} with a summary and outlook.

\vspace{0.1cm}

\begin{figure*}[t]
    \centering
    \includegraphics[width=\textwidth]{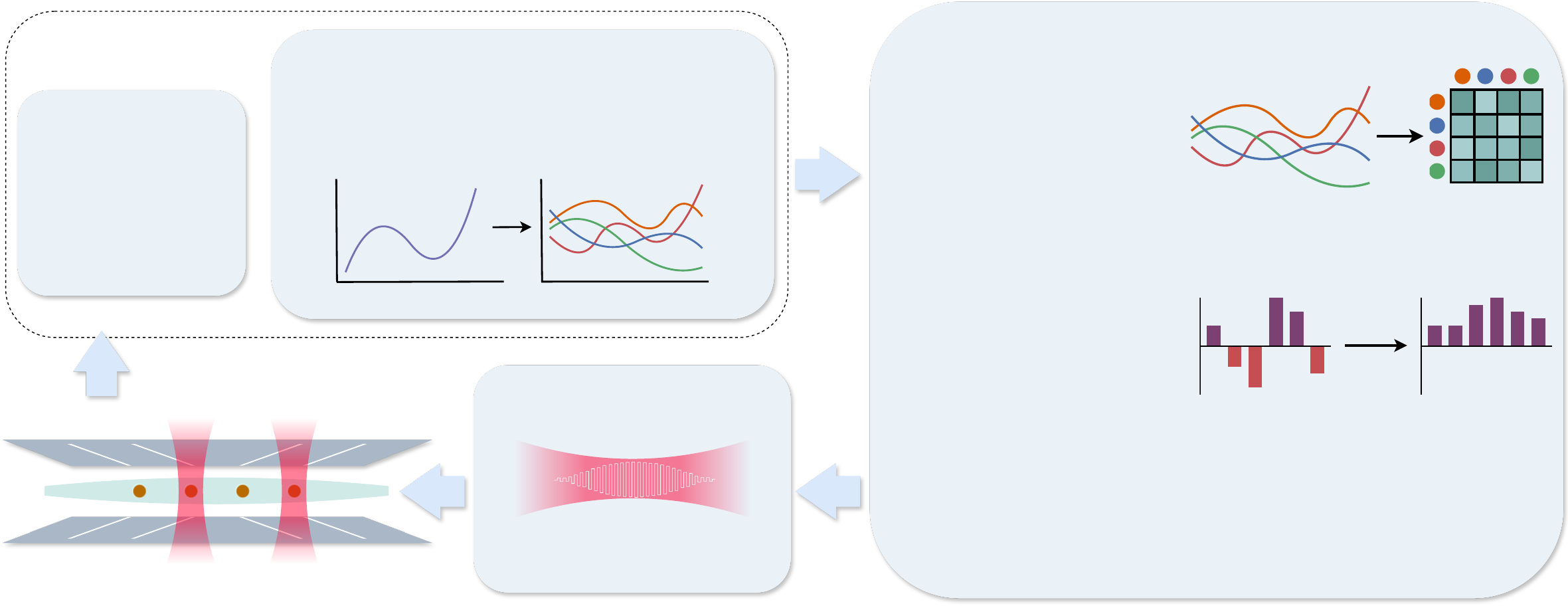}
    \begin{picture}(0,0)
        \put(-226, 86){\ding{172}}
        \put(10, 148){\ding{173}}
        \put(10, 45){\ding{174}}
        \put(-118, 45){\ding{175}}
        \put(-212,155){\makebox(0,0){\textbf{Phonon readout}}}
        \put(-212,135){\makebox(0,0){$\Delta_m,b_j^m,\eta_m$}}
        \put(-83, 185){\makebox(0,0){\textbf{Rabi frequency basis selection}}}
        \put(-83, 168){\makebox(0,0){$\Omega_p(t)=\vec{c}_p\cdot\vec{f}(t)\quad\Omega_q(t)=\vec{c}_q\cdot\vec{f}(t)$}}
        \put(-120, 112){\makebox(0,0){\footnotesize{$t$}}}
        % \put(-152, 133){\makebox(0,0){\rotatebox{90}{\footnotesize{Rabi freq}}}}
        \put(-152, 133){\makebox(0,0){\footnotesize{$\Omega$}}}
        \put(-16, 148){\makebox(0,0){\footnotesize{$f_1(t)$}}}
        % \put(-12, 138){\makebox(0,0){\small{$f_2(t)$}}}
        % \put(-12, 128){\makebox(0,0){\small{$f_3(t)$}}}
        \put(-16, 138){\makebox(0,0){$\vdots$}}
        \put(-16, 122){\makebox(0,0){\footnotesize{$f_4(t)$}}}
        \put(-193, 75){\makebox(0,0){$p$-th}}
        \put(-160, 75){\makebox(0,0){$q$-th}}
        \put(-180, 17){\makebox(0,0){Trapped ion quantum computer}}
        \put(-50, 75){\makebox(0,0){\textbf{Reconstruction}}}
        \put(-48, 30){\makebox(0,0){$\Omega_p(t)\quad\Omega_q(t)$}}
        \put(140, 195){\makebox(0,0){\textbf{Step 1. Matrix representation}}}
        \put(80, 165){\makebox(0,0){\shortstack[c]{$\Theta\to\mathbf{M}\quad E\to\mathbf{H}$ \\\vspace{2pt} \\
        $\vec{\alpha}_j\to\mathbf{A}\quad\tfrac{\partial\vec{\alpha}_j}{\partial\mu}\to\mathbf{A}^{\text{diff}}$}}}
        \put(140, 135){\makebox(0,0){\textbf{Step 2. Positive-extraction approximation}}}
        \put(84, 98){\makebox(0,0){\shortstack[c]{$\mathbf{M}\xrightarrow{\text{flipping}}\tilde{\mathbf{M}}\succ0$ \\
        $\mathbf{H}\xrightarrow{\phantom{\text{flipping}}}\tilde{\mathbf{H}}\quad\;\;$ \\
        $\vec{c}_p,\vec{c}_q\xrightarrow{\phantom{\text{flipping}}}\vec{c}\quad\quad\quad\;\;$}}}
        \put(191, 105){\makebox(0,0){\footnotesize{flipping}}}
        \put(131, 95){\makebox(0,0){\rotatebox{90}{\footnotesize{Eigenvalue}}}}
        % \put(140, 55){\makebox(0,0){\raisebox{0.9ex}{\textbf{Step 3}.} \raisebox{0pt}[2cm][2cm]{\shortstack[c]{
        % \textbf{Quadratically constrained} \\ \textbf{quadratic programming}}}}}
        \put(140, 62){\makebox(0,0){\textbf{Step 3. QCQP optimization}}}
        \put(140, 35){\makebox(0,0){\shortstack[c]{$\text{minimize}\quad \vec{c}^T\tilde{\mathbf{H}}\vec{c}$ \\\vspace{2pt} \\
        $\text{subject to}\quad \vec{c}^T\tilde{\mathbf{M}}\vec{c}=\Theta_{\text{targ}}$}}}
    \end{picture}
    \caption{Framework of heating error suppression.
    \ding{172} Phonon parameters are first characterized experimentally, and a basis $\vec{f}(t) = (f_1(t), f_2(t), \dots, f_L(t))$ for the Rabi frequencies is selected.
    \ding{173} These inputs are fed into our framework to construct a quadratically constrained quadratic programming (QCQP) problem.
     \ding{174} Solving the QCQP yields optimal coefficients $\vec{c}$, which are used to reconstruct the Rabi frequencies using the chosen basis.
     \ding{175} The resulting waveform is then applied to the target qubits to implement a Mølmer–Sørensen gate that is robust to heating errors.}
    % \caption{Framework of heating error suppression. One first determine the basis $f_l(t)$ of the laser pulse and frequency $\mu$. Based on the measured parameters of  the trapped ion systems, control pulse $\Omega_p(t),\Omega_q(t)$ is optimized by our method, and then applied to the trapped ion system.}
    \label{fig:pipeline}
\end{figure*}

\section{PRELIMINARY}
\label{sec:preli}
Mølmer–Sørensen (MS) gate is a widely used entangling gate in trapped-ion quantum computing, owing to its robustness against thermal motion and its ability to generate high-fidelity entanglement between qubits~\cite{molmer1999multiparticle,sorensen1999quantum}. Here, we briefly review the fundamentals of MS gates and outline the criteria for achieving frequency robustness.

Consider a chain of $N$ ions, each possessing a spin-$1/2$ degree of freedom, confined within a trap. When red and blue sideband laser fields, detuned by $\mu$ from the carrier, are applied to the $p$-th and $q$-th ions with Rabi frequencies $\Omega_p(t)$ and $\Omega_q(t)$, respectively, the time evolution of the system is given by~\cite{wu2018noise}
\begin{equation}\label{eq:Utaum}
    U(\tau) = \exp\left(\sum_{j=p,q} \phi_{j}(\tau) \sigma_j + i\Theta(\tau) \sigma_{p} \sigma_{q}\right),
\end{equation}
where
\begin{equation}
    \phi_j(\tau) = \sum_{m=1}^N (\alpha_j^m(\tau) a_m^\dag - \alpha_j^{m*}(\tau) a_m).
\end{equation}
Here, $a_m$ denotes the bosonic annihilation operator acting on the $m$-th motional mode, and $\sigma_j$ represents the Pauli-$X$ operator applied to the $j$-th ion. The coefficient $\alpha_j^m(\tau)$ characterizes the displacement of the $m$-th motional mode and describes its trajectory in phase space. As will be discussed later, these trajectories play a central role in determining heating errors. The second term, $\Theta(\tau)$, represents the accumulated two-qubit rotation angle.
The detailed expressions for $\alpha_j^m(\tau)$ and $\Theta(\tau)$ are provided in Appendix~\ref{app:matexp}.

In the realm of trapped ion quantum computing, constructing a robust quantum gate is crucial for advancing trapped-ion quantum computing, as it serves as a fundamental building block for generating entanglement between qubits. The performance of MS gates is often compromised by various noise sources, such as motional heating, laser intensity fluctuations, and mode frequency instabilities~\cite{ballance2016high,gaebler2016high,baldwin2021high}, which introduce errors that scale unfavorably with the number of ions. As trapped-ion systems expand to accommodate more qubits, mitigating these errors becomes increasingly challenging.

A robust gate must therefore be designed to resist these noise sources without significantly increasing the complexity of the control pulses. Effective error suppression strategies not only improve gate fidelity but also simplify the implementation of error correction protocols, thereby supporting the scalability of trapped-ion quantum processors. Moreover, enhancing gate robustness facilitates the application of fault-tolerant quantum error correction codes, which are essential for achieving practical and large-scale quantum computation. Consequently, the development of robust gates constitutes a critical step toward reliable quantum information processing and broadens the range of quantum algorithms that can be executed on trapped-ion platforms.

A common approach for constructing a robust gate is to employ the frequency-robust Mølmer–Sørensen (MS) gate~\cite{leung2018robust}, which is characterized by a target rotation angle $\Theta_{\text{targ}}$. Specifically, there are three essential criteria for achieving a frequency-robust MS gate:
\begin{enumerate}[(1)]
    \item Ensure that the rotation angle of the quantum gate is adjusted precisely to the target value, i.e. $\Theta(\tau)=\Theta_{\text{targ}}$.
    \item Decouple the spin and phonon degrees of freedom after the gate, i.e. $\alpha_j^m(\tau)=0$.
    \item Ensure that the spin-phonon decoupling remains robust against fluctuations in the mode frequencies or, equivalently, in the laser detuning, i.e. $\partial\alpha_j^m(\tau)/\partial\mu=0$.
\end{enumerate}
These constraints, which are typically regarded as standard requirements, can be satisfied through careful engineering of the Rabi frequencies. In our approach, they remain compatible with our optimization framework, as will be discussed in the next section.

\section{Heating error suppression}
\label{sec:suppression}
%Heating noise constitutes one of the dominate error sources in trapped ion systems~\cite{ballance2016high,gaebler2016high,baldwin2021high}.
Heating noise~\cite{ballance2016high,gaebler2016high,baldwin2021high} in trapped-ion system arises from fluctuating electric fields and laser-induced motional excitations, causing unwanted qubit-phonon entanglement that degrades the precision of entangling gates. This issue becomes more severe as the system size increases~\cite{wu2018noise}, since larger ion crystals exhibit more collective motional modes that interact with environmental noise. Therefore, achieving an MS gate that is robust to heating error is crucial for the realization of large-scale trapped-ion quantum computing, as it ensures high-fidelity operations even in the presence of environmental noise and system imperfections.

For the single-mode case, Refs.~\cite{molmer1999multiparticle,sorensen1999quantum} provide a method to suppress heating errors to arbitrarily low values. By appropriately adjusting the laser detuning $\mu$ and increasing either the Rabi frequency $\Omega$ or the gate time $\tau$, the phase-space trajectory $\alpha(t)$ can be made more localized around the origin while maintaining a fixed entangling phase $\Theta$, thereby reducing heating errors.
% However, this intuitive approach does not directly extend to the multi-mode case. In the presence of multiple phonon modes, the above approach generally cannot ensure that the phase-space trajectories of all modes are simultaneously closed. This is particularly relevant in realistic large-scale trapped-ion systems, where multiple motional modes are inevitably coupled to the laser drive.
However, this intuitive mechanism does not directly extend to the multi-mode case. In multimode systems, as the laser detuning $\mu$ is increased, the entangling phase and the heating error exhibit the same asymptotic scaling with detuning, in contrast to the single-mode case. Consequently, the heating error no longer decreases indefinitely, but instead converges to a finite limit. A detailed derivation is provided in Appendix~\ref{app:detuning_rescaling}.
Therefore, a more refined strategy is required to suppress heating errors in multi-mode systems.

In this section, we present a comprehensive framework for suppressing heating errors in multi-mode trapped-ion quantum computers. The framework is designed to be scalable to systems with an arbitrary number of ions and noise levels, and it supports simultaneous mitigation of additional error sources, such as mode frequency drift.
The key idea is to optimize the control pulses to minimize an analytically derived objective function that quantifies the infidelity induced by heating errors. The framework also allows for a flexible choice of control pulse ansatz, enabling the incorporation of various experimental constraints and requirements. The overall framework is illustrated in Fig.~\ref{fig:pipeline}.

\subsection{heating error estimation}
\label{sec:estimation}
Under the Markov noise approximation, the heating process can be effectively described by the Lindbladian master equation~\cite{Haddadfarshi_2016,he2024scaling}
\begin{equation}
   \frac{\partial\rho}{\partial t}=-i[H(t),\rho]+\sum_{m=1}^N\Gamma_m^\uparrow\mathcal{D}_m^\uparrow\left(\rho\right)+\Gamma_m^\downarrow\mathcal{D}_m^\downarrow\left(\rho\right).
\label{eq:mastereq}
\end{equation}
In the first term, $H(t)$ is the Hamiltonian corresponding to the laser-driven MS gate. %, as described in Eq.~\eqref{eq:Utau}.
% The second term, $\mathcal{D}\left(\rho(t)\right)$, represents the dissipation due to the heating error. Under white noise approximation, the dissipation can be described as
% \begin{equation}
%     \begin{aligned}
%         \mathcal{D}\left(\rho(t)\right)=&\sum_{m=1}^N \Gamma_{m}^{\uparrow}
%         \left(a_m^\dag\rho(t)a_m-\frac{1}{2}\{a_ma_m^\dag,\rho(t)\}\right) \\
%         &+\sum_{m=1}^N \Gamma_{m}^{\downarrow}
%         \left(a_m\rho(t)a_m^\dag-\frac{1}{2}\{a^\dag_ma_m,\rho(t)\}\right),
%     \end{aligned}
%     \label{eq:lind}
% \end{equation}
% where parameters $\Gamma_m^{\uparrow}$ and $\Gamma_m^\downarrow$ referred to as the excitation and relaxation rate respectively.
In the second term, $\mathcal{D}_m^\uparrow$ and $\mathcal{D}_m^\downarrow$ represent the dissipative processes caused by heating-induced excitation and relaxation of the $m$-th motional mode. The parameters $\Gamma_m^{\uparrow}$ and $\Gamma_m^{\downarrow}$ denote the corresponding excitation and relaxation rates.
They signify the average phonon number gain or loss per unit time.
It is assumed that  the heating error is quasi-static, implying that $\Gamma_m^{\uparrow,\downarrow}$ remains constant during the gate operation. These rates can be estimated experimentally~\cite{brownnutt2015ion,bruzewicz2015measurement}, but our error suppression techniques are general and does not rely on the their exact values.

Heating error is an incoherent process, making it particularly challenging to estimate and suppress. Unlike coherent errors, it lacks an associated evolution operator, %like Eq.~\eqref{eq:Utau},
its complicating analytical treatments. Moreover, directly simulating heating error %Eq.~\eqref{eq:mastereq}
 in large-scale ion traps is impractical, as the computational complexity grows exponentially with the number of ions. This presents a major obstacle in developing scalable quantum computing architectures, necessitating efficient estimation techniques and robust error suppression strategies.

To address these challenges, Haddadfarshi and Mintert proposed a method for suppressing heating errors in single-mode systems by employing polychromatic laser pulses~\cite{Haddadfarshi_2016}. They theoretically derived the heating-induced infidelity using a Fourier basis and optimized the control pulses to minimize this error. This approach has been experimentally validated~\cite{shapira2018robust, webb2018resilient}, demonstrating its effectiveness in reducing heating-induced infidelity. However, implementing this method requires multiple laser sources, which makes it experimentally demanding. Moreover, the use of polychromatic lasers is limited to systems with a single motional mode, preventing direct application to larger ion crystals. As the number of ions increases, multiple motional modes become increasingly important, posing a significant challenge for scaling up trapped-ion quantum computers~\cite{PhysRevX.14.041017, shapira2023fastdesignscalingmultiqubit}. Therefore, developing more scalable and flexible techniques to suppress heating errors is crucial for advancing fault-tolerant quantum computing with trapped ions.

In this work, we adopt a recent result that provides an efficiently computable upper bound on the leading-order contribution of heating-induced infidelity~\cite{he2024scaling}, which serves as an effective cost function for optimization. Let $\rho(\tau)$ be the final state under the noise %evolution of Eq.~\eqref{eq:mastereq}
and $\rho_{\text{ideal}}$ be the corresponding ideal state (by setting $\Gamma_m^{\uparrow,\downarrow}=0$). Ref.~\cite{he2024scaling} show that when the heating noise is small, the infidelity can be accurately estimated according to the trajectories in phase spaces.
More specifically, when $\Gamma_m^{\uparrow,\downarrow}$ are sufficiently small, and the initial state of spin and phonon degree of freedoms are separable, the infidelity satisfies $1-F(\tr_{\text{ph}}\rho(\tau),\tr_{\text{ph}}\rho_{\text{ideal}})\lesssim E$, where

\begin{equation}
    \begin{aligned}
        E&=\sum_{j_1,j_2=p,q}\left|E_{j_1,j_2}\right|,\\
        E_{j_1,j_2}&=
        \sum_{m=1}^N\left(\Gamma_m^\uparrow+\Gamma_m^\downarrow\right)
        \int_0^\tau \mathrm{d}t \alpha_{j_1}^{m*}(t)\alpha_{j_2}^m(t).
    \end{aligned}
    \label{eq:bound}
\end{equation}
Here, $\tr_{\text{ph}}$ denotes the partial trace operation over the phonon states, leaving only the spin state. Fortunately, $E$ is efficiently computable as oppose to the exact infidelity, because it has eliminated higher-order terms. Moreover, numerical results in Appendix~\ref{app:infid&l2norm} have indicated that it serves as a well-estimation of the actual infidelity, especially when the Rabi frequencies are reasonable. Therefore, $E$ is a suitable cost function for further heating error suppression.

\subsection{positive-extraction approximation}
\label{sec:pe_approx}
The cost function $E$, along with three constraints for achieving a frequency-robust MS gate, defines the optimization problem for suppressing heating errors. However, solving this problem directly is challenging due to its inherent non-linearity and non-convexity. First, the objective function~\eqref{eq:bound} involves a sum of absolute values, breaking the linearity of the optimization landscape. Second, the rotation angle $\Theta(\tau)$ is generally not a positive semidefinite function of the control pulse $\Omega_j(t)$, which further complicates the optimization. Together, these challenges render the optimization problem NP-hard.

%  Various approaches have been explored for achieving robustness in quantum gate operations, including frequency modulation~\cite{leung2018robust}, phase modulation~\cite{milne2020phase, liu2023efficient}, and amplitude modulation~\cite{zhu2006arbitrary}. While each approach offers its unique advantages, we focus on amplitude modulation, although the generalization to other methods is possible.

To make the solution efficiently, we propose the positive-extraction approximation to simplify the optimization. With this approach, the optimization is reduced to a positive-semidefinite quadratically constrained quadratic programming (QCQP) problem, which can be tackled using standard optimization techniques. The approach consists of two main steps. The first step is to observe that the cross terms $E_{p,q}$ and $E_{q,p}$ in Eq.~\eqref{eq:bound} can be safely neglected, as they are bounded by the diagonal terms. This approximation removes the non-linearity of the objective function, leaving only positive diagonal terms. In the second step, we adjust the rotation angle $\Theta(\tau)$ to be positive semidefinite by applying individually designed Rabi frequencies $\Omega_p(t)$ and $\Omega_q(t)$. Specifically, we decompose the Rabi frequencies into a set of basis components and assign opposite signs to the coefficients of $\Omega_p(t)$ and $\Omega_q(t)$ for those components that would otherwise yield a negative contribution to the rotation angle.

To obtain a concrete and general formulation, we introduce a set of linearly independent basis functions $\vec{f}(t) = (f_1(t), f_2(t), \dots, f_L(t))^T$, so that the time-dependent Rabi frequencies can be written as $\Omega_j(t) = \vec{c}_j \cdot \vec{f}(t)$, where $\vec{c}_j$ is the coefficient vector for ion $j$. With this representation, the heating error cost function $E$ in Eq.~\eqref{eq:bound} can be expressed in matrix form as
\begin{align}
E_{j_1,j_2} = \vec{c}_{j_1}^T \mathbf{H}^{(j_1,j_2)} \vec{c}_{j_2}, \label{eq:se}
\end{align}
where the matrix $\mathbf{H}^{(j_1,j_2)}$ is the matrix representation of $E_{j_1,j_2}$ that encapsulates the contributions from all motional modes and is defined in Appendix~\ref{app:matexp}.
and the optimization problem becomes
\begin{subequations}\label{eq:m1}
    \begin{align}
        \minimize \quad &\sum_{j_1,j_2=p,q}\left|E_{j_1,j_2}\right|,\label{eq:s} \\
    \text{subject to}\quad &\vec{c}_{p}^T\mathbf{M}\vec{c}_{q}=\Theta_{\text{targ}},\label{eq:sa}\\
    &\mathbf{A}\vec{c}_j=\vec{0},\label{eq:sb} \\
    &\mathbf{A}^{\text{diff}}\vec{c}_j=\vec{0},\label{eq:sc}
    \end{align}
\end{subequations}
where Eqs.~\eqref{eq:sa}–\eqref{eq:sc} correspond to the three criteria for frequency-robust MS gates (see Section~\ref{sec:preli}). The matrices $\mathbf{M}$, $\mathbf{A}$, and $\mathbf{A}^{\text{diff}}$ are functions of the chosen pulse basis ${f_l(t)}$, and represent the matrix forms of $\Theta(\tau)$, the phonon displacements $\vec{\alpha}_j = (\alpha_j^1(\tau), \dots, \alpha_j^N(\tau))^T$, and their detuning derivatives $\partial \vec{\alpha}_j/\partial\mu$, respectively. Their explicit forms are given in Appendix~\ref{app:matexp}.

The optimization problem defined in Eq.~\eqref{eq:m1} is NP-hard, primarily  due to the non-linearity of the objective function and the non-convexity of the constraints. To solve this problem efficiently, we introduce two theorems that simplify the structure of the problem.

The first theorem allows us to eliminate the absolute value terms in the objective function, thereby restoring the linear structure needed for tractable optimization.
\begin{restatable}{theorem}{thmone}
    For arbitrary trajectories $\alpha_j^m(t)$ in phase space, the cross-terms in the heating error are bounded by the diagonal terms:
    \begin{equation}
        |E_{p,q}|+|E_{q,p}|\leq|E_{p,p}|+|E_{q,q}|.
    \end{equation}
    \label{thm:thm1}
\end{restatable}
\noindent The proof of Theorem~\ref{thm:thm1} follows directly from Cauchy--Schwarz and the arithmetic--geometric mean (GM-AM) inequality.

By applying Theorem~\ref{thm:thm1}, the objective function in Eq.~\eqref{eq:s} can be significantly simplified. Since $E_{p,p}$ and $E_{q,q}$ are positive by construction, we estimate the heating error by
\begin{equation}
\sum_{j_1,j_2=p,q} |E_{j_1,j_2}| \sim E_{p,p} + E_{q,q},
\label{eq:estimation}
\end{equation}
which removes the absolute values and restores linearity. This approximation not only simplifies the problem but also highlights the advantage of adopting a matrix-based formulation.

The second theorem ensures that the matrix involved in the optimization is positive semidefinite. After applying Theorem~\ref{thm:thm1}, the original optimization problem in Eq.~\eqref{eq:m1} reduces to a quadratically constrained quadratic programming (QCQP) problem.
Although this QCQP is generally NP-hard due to the non-positive-definiteness of the matrix $\mathbf{M}$ in Eq.~\eqref{eq:sa}, this difficulty can be circumvented by noting thatthe left and right multipliers of $\mathbf{M}$ are two distinct vectors. Specifically, we introduce the following theorem:

\begin{restatable}{theorem}{thmtwo}
    There exists matrices $\mathbf{S}_p$, $\mathbf{S}_q$ and a positive-semidefinite matrix $\mathbf{\tilde M}$ such that, for any $\vec{c}$ satisfying $\vec{c}^{T} \mathbf{\tilde M}\vec{c}=\Theta_{\text{targ}}$, Eq.~\eqref{eq:sa}, Eq.~\eqref{eq:sb} and \eqref{eq:sc} can be simultaneously satisfied by setting $\vec{c}_p=\mathbf{S}_p\vec{c}$ and $\vec{c}_q=\mathbf{S}_q\vec{c}$.
    \label{thm:thm2}
\end{restatable}

\noindent The construction of $\mathbf{\tilde M}$ involves projecting $\mathbf{M}$ onto the kernel space defined by Eq.\eqref{eq:sb} and Eq.\eqref{eq:sc}, followed by flipping the signs of its eigenvalues while preserving the eigenvectors to ensure positive semidefiniteness. The explicit forms of $\mathbf{\tilde M}$, $\mathbf{S}_p$, and $\mathbf{S}_q$ are provided in the proof in Appendix~\ref{app:proof}.

With Theorem~\ref{thm:thm2}, the original problem can now be reformulated as the following positive-semidefinite QCQP:
\begin{equation}
    \begin{aligned}
        &\minimize\quad \vec{c}^T\mathbf{\tilde H}\vec{c},\\
        &\text{subject to}\quad\vec{c}^T\mathbf{\tilde M}\vec{c}=\Theta_{\text{targ}},
    \end{aligned}
    \label{eq:optprob}
\end{equation}
where $\mathbf{\tilde H}=\mathbf{S}_p^T\mathbf{H}^{(p,p)}\mathbf{S}_p+\mathbf{S}_q^T\mathbf{H}^{(q,q)}\mathbf{S}_q$. This optimization can be efficiently solved via the method of Lagrange multipliers. Once the optimal solution $\vec{c}$ is obtained, the corresponding control pulses are reconstructed by $\vec{c}_p = \mathbf{S}_p \vec{c}$ and $\vec{c}_q = \mathbf{S}_q \vec{c}$.

Our approach is general and applicable to any ion number, any choice of control pulse basis, any laser detuning, and any sufficiently small noise level. Furthermore, it removes the need for polychromatic lasers, improving experimental feasibility. Nevertheless, our method remains compatible with previous polychromatic approaches~\cite{Haddadfarshi_2016} when the Fourier basis is chosen, as discussed in Appendix~\ref{app:premethod}. In the following, we demonstrate the effectiveness of our method through numerical simulations and comparisons with several baseline methods.

\section{numerical results}
\label{sec:results}
We consider the piecewise constant waveforms as follows
\begin{equation}
    f_l(t)=\begin{cases}
        1 \quad \frac{l-1}{L}\tau\leq t<\frac{l}{L}\tau \\
        0 \quad \text{otherwise}
    \end{cases},
    \label{eq:piecewisebasis}
\end{equation}
which is the simplest waveform ansatz and easy to implement in experiment. To demonstrate the generality, we also provide an example for Fourier basis in Appendix~\ref{app:premethod}. In the main text, however, we focus on piecewise waveform only.

\subsection{Infidelity Comparison with Existing Methods}
To demonstrate the improvement of our method, we use two baseline waveform for comparison.
% The first one is the method in~\cite{leung2018robust}, \red{\sout{which only ensures the vanishes and robustness of $\alpha_j^m(\tau)$, and}} devoid of any heating error minimization.
The first one corresponds to the approach of Ref.~\cite{leung2018robust}, which constructs waveforms by enforcing the required gate constraints only, without optimizing any cost function related to heating errors.
We denote this method as the \textit{conventional} method.
Besides, one might argue that the heating error is closely related to the laser power due to the quadratic dependency in Eq.~\eqref{eq:bound}. In the second baseline waveform, denote as the \textit{min-Rabi} method, we minimizes the $L^2$-norm of the Rabi frequency $\int_0^\tau |\Omega(t)|^2 dt$ instead of the cost function in Eq.~\eqref{eq:bound}.

For any given ion number $N$ and laser detuning $\mu$, we segment the waveform into $L=6N+20$ sections. Subsequently, we apply these three methods: our method Eq.~\eqref{eq:optprob}, the \textit{conventional} method, and the \textit{min-Rabi} method, yielding three optimized waveforms. For small ion numbers ($N\leq6$), a numerical simulation of the system is conducted, obtaining the infidelities of all three waveforms. Throughout our numerical simulation, we set the cut-off phonon number as $N_{\text{cut}}=10$. For larger ion numbers ($N>6$), numerical calculation of the infidelity becomes challenging due to the exponential increase of the Hilbert space dimension. In such cases, we use the upper bound of infidelity after the positive-extraction approximation given in Eq.~\eqref{eq:estimation} as an error estimation. In Appendix~\ref{app:infid&l2norm}, we provide a detailed comparison between this error estimation and the actual infidelities, demonstrating that the estimation is highly accurate for predicting heating error. This confirms that our proposed error estimation method reliably captures the key features of the heating error, making it an effective tool for evaluating and optimizing performance.

\begin{figure*}[t]
    \centering
    \includegraphics[width=\linewidth]{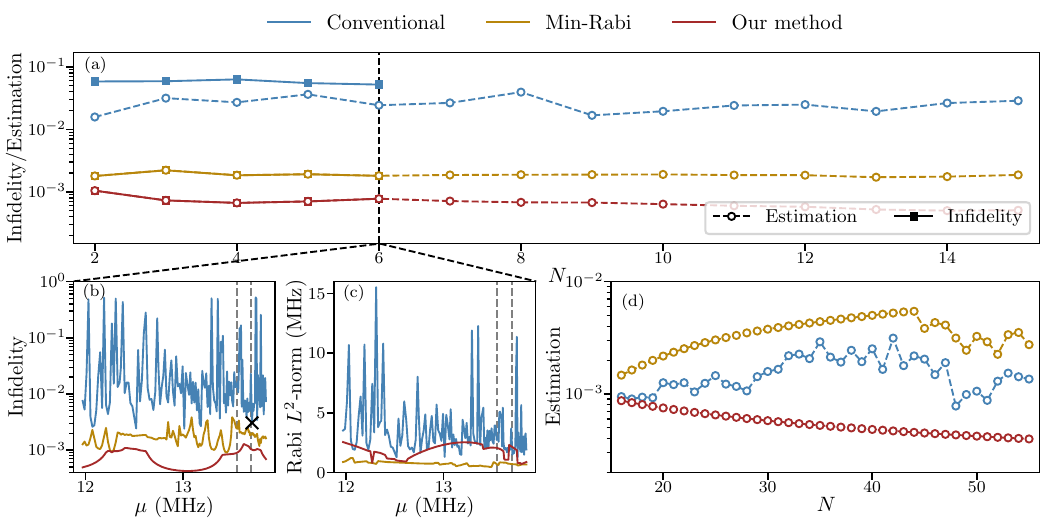}
    \caption{Infidelities, estimations and Rabi $L^2$-norms optimized by three methods. (a) and (d): average infidelity and estimation versus numbers of ions $N$. The dashed lines represent error estimations calculated using Eq.~\eqref{eq:estimation} and the solid lines represent the simulated infidelities. We set $\Gamma_m^\uparrow=\Gamma_m^\downarrow=100\text{ phonon}/s$ and gate duration $\tau=150\mu s$. Only the left-most two ions are involved, i.e. $p=1$ and $q=2$. For each $N$, $20(N+1)$ laser detunings $\mu$ evenly distributed among $N$ modes are choosen to be optimized. In subfigure (a), the trap is weak and can only hold up to 15 ions, whereas the trap in subfigure (d) is stronger and can hold up to 55 ions. (b) and (c): detailed infidelities and L2-norms of Rabi frequencies versus laser detunings $\mu$ for $N=6$ in the subfigure (a). The two dashed vertical lines locate the two highest mode frequencies, i.e. the Center-of-Mass (COM) mode and the breathing mode. In all subfigures, the blue line represents the conventional baseline waveforms without any heating error optimization; the yellow line represents the min-Rabi baseline waveforms optimized for $L^2$-norm of Rabi frequencies; the red line represents waveforms generated by our method.}
    \label{fig:avg_infid}
\end{figure*}

In Fig.~\ref{fig:avg_infid}(a), we demonstrate the average infidelities and error estimations over different laser detunings $\mu$ in a weak trap. For any ion numbers that can be trapped, large improvement rates relative to the baseline methods are observed. In particular, the average infidelity of our methed are only $6.67\times10^{-4}$ at $N=4$, while that of the conventional method reaches $6.31\times10^{-2}$. The improvement rate compare to the conventional methods is $94.7$. Even comparing to the min-Rabi waveform (with infidelity $1.84\times10^{-3}$ at $N=4$), our method still provides $2.77$ times improvement. These results verify that our method catches the fundamental mechanism of heating noise dissipation, and outperforms the simple Rabi strength minimization strategy.

For systems with a larger ion number ($N > 15$), the trap used in Fig.~\ref{fig:avg_infid}(a) can no longer trap the ions, necessitating the use of a stronger trap. Fig.~\ref{fig:avg_infid}(d) shows the average error estimations in this stronger trap. %\red{While the error estimations of all three methods are similar at small $N$, this regime is not representative of typical operating conditions. With a small ion number, the large inter-ion spacing leads to weak Coulomb interactions, making this parameter choice less realistic for typical trapped-ion systems.}
% While the error estimations of all three methods are similar at $N=16$, they diverge as $N$ increases.
As $N$ increases, our method consistently shows a steady and smooth decrease in heating error. This behavior is likely attributable to the growing number of optimization degrees of freedom with increasing $N$. At $N = 44$, our method results in a heating error estimation of only $4.56 \times 10^{-4}$, a much smaller value compared to the conventional method ($2.18 \times 10^{-3}$) and the min-Rabi method ($5.44 \times 10^{-3}$), corresponding to a fivefold reduction. Yet, we note that the actual improvements, as shown in Fig.~\ref{fig:avg_infid}(d), are parameter dependent. In particular, for small $N$, the large inter-ion spacing leads to weak interactions, so although the improvement in this regime is not large, it is less representative of typical operating conditions. The trend with respect to $N$ is particularly significant because it demonstrates that our approach remains robust even as the system size grows. On the other hand, both the conventional and min-Rabi methods exhibit an increase in heating error as $N$ increases, due to the larger number of heating channels, suggesting that these methods become less effective with larger systems.

We note that our simulation is subject to numerical errors due to the limited cut-off phonon number $N_{\text{cut}}$. This error is larger for the conventional methods because its average phonon number during the operation is higher. There are three reasons keeping us from increasing $N_{\text{cut}}$ to get a converge result. First, increasing $N_{\text{cut}}$ further is challenging because of the increase of the Hilbert space dimension. Second, the trend of current result is qualitatively correct and sufficient for our discussion. Third, these numerical errors reflect practical issues, which will be discussed later.

% \begin{figure}[t]
%     \centering
%     \includegraphics[width=\linewidth]{fig/infid&amp.pdf}
%     \caption{Infidelity and L2 norm of laser power versus laser frequency. (a) and (b): infidelity for $N=2$ and $N=6$ respectively; (c) and (d): L2 norm of laser power for $N=2$ and $N=6$ respectively. In all subfigures, the purple line represents the conventional baseline waveforms without any heating error optimization; the blue line represents the min-power baseline waveforms optimized for optimal L2 norm; the red line represents waveforms generated by our method Eq.~\eqref{eq:optprob}. Dashed orange vertical lines indicate the positions of phonon mode frequencies, dividing the laser frequency into $N+1$ intervals. Each interval is further divided into 20 points, and the system is optimized and simulated at these points with $p=1$, $q=2$, $\Gamma_m^\uparrow=\Gamma_m^\downarrow=100\text{ phonon/s}$ and $\tau=150\mu s$. The black crosses in (b) and (d) indicate a special laser frequency utilized in Fig.~\ref{fig:alpha}}.
%     \label{fig:infid&amp}
% \end{figure}

\subsection{Relations between Rabi strength and heating error}
To better understand these issues and the relation between Rabi strength and heating error, we demonstrate the infidelities and $L^2$-norms of Rabi frequencies versus $\mu$ for different methods in Fig.~\ref{fig:avg_infid}(c) and (d). We choose $N=6$ as a typical example, while the performance for other ion numbers are provided in Appendix~\ref{app:infid&l2norm}. The infidelity of the conventional method exhibits sharp fluctuations. For example, for $N=6$, the infidelity oscillates significantly, with the highest recorded infidelity reaching 0.523~\footnote{This value might subject to numerical error due to the high average phonon excitation} at a detuning of 13.741 MHz. Except for this peak, the average infidelity remains relatively low at $5.21\times10^{-2}$.

The sharp fluctuation in infidelity can be attributed to the variation of Rabi $L^2$-norms. As illustrated in Fig.~\ref{fig:avg_infid}(c) and (d), the peak of infidelity coincide with the peak of Rabi $L^2$-norms. Specifically, for $N=6$, the highest infidelity is achieved at a high $L^2$-norm of $8.51$ MHz, contrasting with the typical strength of only $1$ MHz. This observation is consistent with the expression of $U(t)$: %Eq.~\eqref{eq:alpha}:
as the Rabi strength $\left|\Omega_j(t)\right|$ increases, the phase space displacement $\alpha_j^m(t)\propto\Omega_j(t)$ also increase, and hence the effect of heating error. We also note that except for the vulnerability to heating noise, high Rabi strength also introduces other problems in practice. Firstly, the control pulse may goes out of the region where Lamb-Dicke approximation is applicable. %, rendering the applicability of Hamiltonian in Eq.~\eqref{eq:H}.
Secondly, laser with higher strength will also have a larger control error in experiment. Thus, the numercial errors caused by limited cut-off $N_\text{cut}$ are kept to reflect these practical issues.

On the other hand, merely reducing the Rabi $L^2$-norm is insufficient for mitigate heating errors. For min-Rabi method, while the fluctuation assuages notably, the infidelities do not reach their minimum. In contrast, our method has the lowest infidelity across all $\mu$. This also indicates that the Eq.~\eqref{eq:bound} is an appropriate cost function for heating error suppression. At the same time, the Rabi $L^2$-norms of our method remain stable, making it suitable for practical implementation across a spectrum of laser detunings $\mu$. Such a stability can be attributed to the positive definiteness of the matrix $\mathbf{\tilde{M}}$, which prevents the cancellation of different components during the rotation by an angle $\Theta_{\text{ideal}}$.

\begin{figure}[t]
    \centering
    \includegraphics[width=\linewidth]{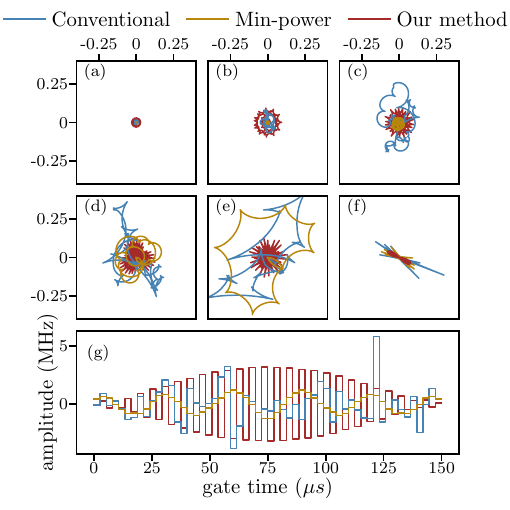}
    \caption{
    %Specific parameters during the gate duration for the first ion $j=1$ within a chain of $N=6$ ions at $\mu=13.703\text{MHz}$, as marked by a cross-reference in Fig.~\ref{fig:avg_infid}(b).
    (a)-(f):  Trajectory of $\alpha_1^m(t)$ in phase space for ion number $N=6$ and $\mu=13.703$ MHz as marked in Fig.~\ref{fig:avg_infid}(b). (a) to (f) correspond to $m=1,2,\dots,6$ respectively, and in particular, (e) is the breathing mode, and (f) is the COM mode. (g): The optimized Rabi frequeny $\Omega_1(t)$ on the first ion as a function of time.
    In all subfigures, the blue line represents the conventional baseline waveforms without any heating error optimization; the yellow line represents the min-Rabi baseline waveforms optimized for $L^2$-norm of Rabi frequencies; the red line represents waveforms generated by our method.}
    \label{fig:alpha}
\end{figure}

\subsection{trajectories in phase space and optimized waveforms}
Eq.~\eqref{eq:bound} suggests that the heating error can be estimated by considering the average modulus length of $\alpha_j^m(t)$. To provide a more detailed insight into the effect of the algorithm, we plot the trajectory of $\alpha_j^m(t)$ in phase space in Fig.~\ref{fig:alpha}. We focus on one of the entangled ions within a chain of six ions and set the laser detuning to $\mu=13.704$ MHz, as marked by a cross-reference in Fig.~\ref{fig:avg_infid}(b), which is close to the center-of-mass (COM) mode. It becomes evident from the figure that both the min-Rabi method and our proposed method effectively compress the trajectory of $\alpha_j^m(t)$ near the origin. Notably, for the first three modes in Fig.~\ref{fig:alpha}(a), (b) and (c), the min-Rabi method achieves a slightly compression than ours, owing to its utilization of very small Rabi strength. However, when considering the modes closer to the laser detuning in Fig.~\ref{fig:alpha}(d) and (e), simple Rabi strength optimization fails to yield substantial reduction, whereas our method exhibits superior control, resulting in trajectories clustered around the origin, thereby minimizing heating errors.

In addition to heating errors, violations of the Lamb–Dicke approximation constitute another important error source associated with the magnitude of the motional phase-space displacement. Large displacements correspond to higher average phonon occupation during the gate operation, which may violate the Lamb–Dicke condition $\eta_m^2(\bar{n}+1)\ll1$. Here, the average phonon number can be estimated as
\begin{equation}
    \bar{n}\sim \sum_j \sum_m \left|\alpha_j^m\right|^2 .
\end{equation}
Consequently, optimizing the average displacements not only suppresses heating-induced errors but also mitigates errors arising from violations of the Lamb–Dicke approximation.

Fig.~\ref{fig:alpha}(g) shows gate waveforms optimized by three methods. For min-Rabi waveform, the highest Rabi strength is 1.268 MHz, which is much lower than other two waveforms, reaching 5.801 MHz for conventional method and 3.204 MHz for our method. Nevertheless, our method still generates waveforms within a reasonable strength range.
% \red{\sout{If an even lower Rabi frequency is desired, we note that alternative waveform bases can be employed. In particular, as shown in the Appendix~\ref{app:premethod}, using a Fourier basis can significantly reduce the Rabi frequency while maintaining effective suppression of heating errors.}}
% In addition, our method offers another advantage over the other two: the strength of the generated waveforms varies smoothly, which is advantageous for experimental implementation. Although the waveform consists of both positive and negative phases, it can be experimentally implemented with ease by simply flipping the phase. While our method does not explicitly enforce this behavior, it is observed to occur in the majority of cases.

Interestingly, the optimized waveform produced by our numerical procedure exhibits two qualitative features that resemble strategies previously used to suppress detuning-related errors: a slowly varying envelope and alternating sign segments~\cite{hayes2012coherent,tyler2024laser}.
% Similar structures have been discussed in earlier works in the context of reducing residual motional displacement.
Yet, we emphasize that our optimization does not explicitly target for detuning-error suppression. Instead, the waveform is obtained by minimizing the error induced by heating noise, and the observed structures arise naturally from this optimization.

Such waveforms are advantageous in practice: they are easier to implement experimentally, and they can also help suppress residual motional displacement, as discussed in Sec.~\ref{sec:detuning}. One possible concern is that the sign alternation may be too rapid for experimental implementation, especially since the number of segments $L=6N+20$ increases with system size. However, this requirement is mainly due to the fact that our optimization includes all phonon modes. In practice, modes that are far detuned from the laser (e.g., Fig.\ref{fig:alpha}(a) and (b)) contribute negligibly and can be safely neglected, which significantly reduces the required number of segments. Alternatively, one may adopt smoother basis functions, such as a Fourier basis, to parameterize the waveform; we provide a demonstration of this approach in Appendix~\ref{app:premethod}

\subsection{Robustness Against Heating Rate Mismatch}
While most system parameters can typically be characterized with high precision, the excitation and relaxation rates $\Gamma_{m}^{\uparrow,\downarrow}$ often involve greater uncertainty.
The cost function in Eq.~\eqref{eq:bound} explicitly depends on these rates, suggesting that accurate knowledge of $\Gamma_{m}^{\uparrow,\downarrow}$ may be required for the optimization to be valid. Nevertheless, we find that our method remains effective even when these parameters vary, indicating that a rough estimation of $\Gamma_{m}^{\uparrow,\downarrow}$ is sufficient in practice.

\begin{figure}[t]
    \centering
    \includegraphics[width=\columnwidth]{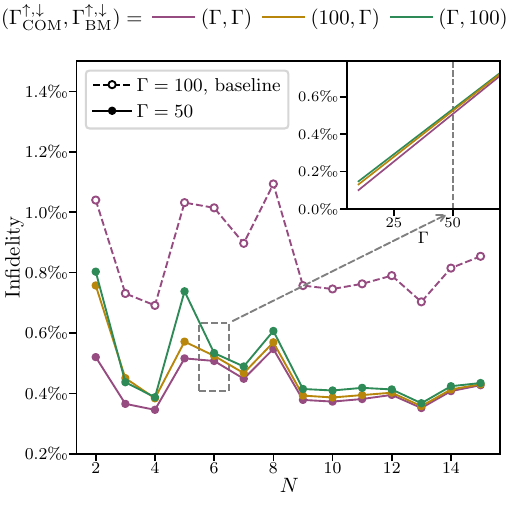}
    \caption{Infidelity of our optimal waveform under various excitation and relaxation rates. Optimal waveforms are optimized using $\Gamma_m^{\uparrow,\downarrow} = 100$ phonon/s and a laser detuning $\mu$ near the COM mode. Three scenarios are considered to test robustness for varying the excitation and relaxation rates $\Gamma_m^{\uparrow,\downarrow}$ using the parameter $\Gamma$: Scenario 1 (purple line) where all motional modes have rates set to $\Gamma$; Scenario 2 (yellow line) where all modes except the COM mode are set to $\Gamma$ while $\Gamma_{\text{COM}} = 100$ phonon/s; and Scenario 3 (green line) where all modes except the breathing mode are set to $\Gamma$ while $\Gamma_{\text{BM}} = 100$ phonon/s. The main panel shows the infidelity versus ion number $N$, comparing the optimized baseline ($\Gamma = 100$ phonon/s, dashed line) with mismatched values ($\Gamma = 50$ phonon/s, solid lines). The inset shows infidelity at fixed $N=6$ as $\Gamma$ varies from 10 to 75 phonon/s in all three scenarios.}
    % \caption{Infidelity of our optimal waveform under various excitation and relaxation rates.  Optimal waveforms are optimized using $\Gamma_m^{\uparrow,\downarrow}=100\text{ phonon/s}$ and a laser detuning $\mu$ near the COM mode. Excitation and relaxation rates for modes other than the COM and breathing modes are varied using parameter $\Gamma$. Three scenarios are considered: Scenario 1 (purple line) where both COM and breathing modes vary with $\Gamma$, Scenario 2 (yellow line) where only the breathing mode varies with $\Gamma$ while COM mode $\Gamma_{\text{COM}}=100\text{ phonon/s}$, and Scenario 3 (green line) where only the COM mode varies with $\Gamma$ while breathing mode $\Gamma_{\text{BM}}=100\text{ phonon/s}$. The main figure shows infidelities versus ion number $N$. The dashed line represents the baseline with $\Gamma=100\text{ phonon/s}$, the setting used for optimization, while solid lines denote $\Gamma=50\text{ phonon/s}$. In the inset, ion number is fixed at $N=6$, with $\Gamma$ varied from 10 to 75 $\text{ phonon/s}$ for all three scenarios.}
    \label{fig:Gamma}
\end{figure}

\begin{figure*}[t]
    \centering
    \includegraphics[width=\linewidth]{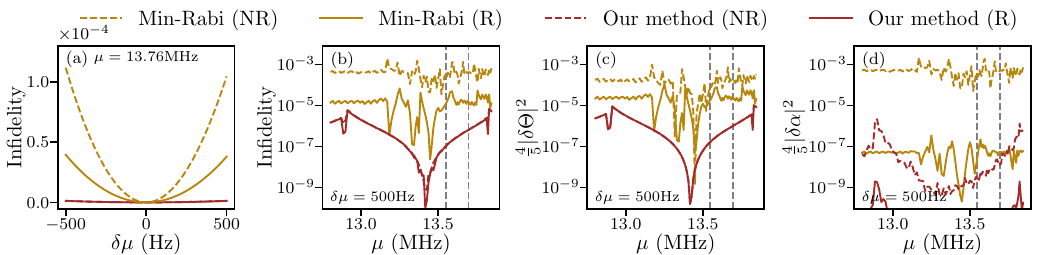}
    \caption{Infidelity and fluctuations versus detuning error $\delta \mu$ for the left-most two ions among $N=4$ ions. (a): Infidelity under various detuning errors $\delta \mu$ at a nominal detuning of $\mu = 13.76\,\text{MHz}$. (b)-(d): Infidelities, rotation-angle fluctuations $|\delta\Theta|^2$, and residual displacements $|\delta\alpha|^{2}$ versus various detuning $\mu$ under a fixed detuning error of $\delta \mu=500\,\text{Hz}$. In all subfigures, yellow curves correspond to the min-Rabi baseline and red curves to our method; NR denotes the non-robust versions (dashed), and R denotes the robust versions (solid) incorporating $\alpha$-robustness constraints. The two dashed vertical lines locate the two highest mode frequencies, i.e. the COM mode and the breathing mode. Here, $\delta\Theta$ denotes the deviation of the accumulated rotation angle from its target value, and $|\delta\alpha|^{2}$ denotes the total residual motional displacement $\sum_{j,m}|\alpha_j^{m}(\tau)|^{2}$. In (a)-(c), the NR and R versions of our method (dashed and solid red lines) nearly overlap, indicating that our method inherently provides robustness against detuning errors even without explicit constraints.}
    \label{fig:detuning}
\end{figure*}

In Fig.~\ref{fig:Gamma}, we optimize the waveform using our algorithm with parameters $\Gamma_m^{\uparrow,\downarrow}=100$ phonon/s, then simulate the infidelity under varied parameters. The laser detuning is set near the COM mode (the mode with the highest frequency) with a offset of $\delta=2\pi\text{kHz}$. We consider three scenarios. The first scenario is where all phonon modes share identical excitation and relaxation rates $\Gamma_m^{\uparrow,\downarrow} = \Gamma < 100$ phonon/s The second scenario is where the COM mode rates are fixed at 100 phonon/s while other modes have $\Gamma_{m \neq \text{COM}}^{\uparrow,\downarrow} = \Gamma < 100$ phonon/s;. The third scenario is where he breathing mode rates are fixed at 100 phonon/s while other modes have $\Gamma_{m \neq \text{BM}}^{\uparrow,\downarrow} = \Gamma < 100$ phonon/s.
In all scenarios, infidelities decrease approximately linearly with $\Gamma$, as shown in the inset of Fig.~\ref{fig:Gamma}. These results indicate that waveforms optimized for a specific parameter remain effective under variations of $\Gamma$.

An interesting observation is that although $\mu$ is close to the COM mode, the impact of changes in $\Gamma_{\text{COM}}^{\uparrow,\downarrow}$ is smaller than that of $\Gamma_{\text{BM}}^{\uparrow,\downarrow}$. This means that the contribution of heating error for COM mode is not dominant in our algorithm, which can also be observed from Fig.~\ref{fig:alpha}(e) and (f). As an example, for $N=5$, increasing $\Gamma_{\text{BM}}^{\uparrow,\downarrow}$ from $50\text{ phonon/}s$ to $100\text{ phonon/}s$ results in a 43.0\% increase in infidelity, while increasing $\Gamma_{\text{COM}}^{\uparrow,\downarrow}$ by the same amount only causes a $10.8\%$ increase. This discrepancy can be understood from the pulse shape in Fig.~\ref{fig:alpha}(g). For conventional methods, the pulse shape and hence the trajectory is unordered. Because $\mu$ is close to the COM mode, this makes the trajectory move away from the original point rapidly. On the contrary, the pulse optimized by our algorithm is keeping flipping around x-axis. Accordingly, the trajectory in phase space oscillates around the original point, because it turns the direction of moving frequently with appropriate angles.

\subsection{Robustness Against Detuning Error}
\label{sec:detuning}
In addition to heating noise, fluctuations in the laser detuning $\mu$, denoted as $\delta\mu$, constitute another important error source. According to Eq.~\eqref{eq:Utaum}, $\delta\mu$ affects both residual displacements $\alpha_j^m(\tau)$ and accumulated rotations angle $\Theta(\tau)$.  The former can be suppressed by imposing additional robustness constrains~\cite{leung2018robust}. Our framework is fully compatible with this technique, allowing the suppression $\alpha_j^m(\tau)$ be incorporated alongside heating-error optimization [Fig.~\ref{fig:detuning}(d)].
%have shown that detuning-induced errors can be mitigated by imposing additional robustness constraints that explicitly suppress residual displacements $\alpha_j^m(\tau)$~\cite{leung2018robust}. Our framework is fully compatible with this technique, allowing such displacement-robust constraints to be incorporated alongside heating-error optimization [Fig.~\ref{fig:detuning}(c)].}
More importantly, our method optimized for heating error provides \textit{additional suppression} on the fluctuations of $\Theta(\tau)$, which is equally essential. %This suppression is at least equally important to the residual displacements.
%as a byproduct of the heating-error–based optimization.

To demonstrate the robustness, we consider a system of $N=4$ ions and optimize the waveform for the leftmost pair ($p=1$ and $q=2$) with methods in Sec.~\ref{sec:suppression}. The gate duration is fixed to $\tau = 150\mu\text{s}$, and the control waveform is discretized into $L = 6N + 20 = 44$ piecewise-constant segments. We benchmark our method against the min-Rabi baseline. For both schemes, we consider two variants: incorporation with the robustness of $\alpha_j^m$ (method in Ref.~\cite{leung2018robust}, denoted as R) or not (denoted as NR). %The $\alpha_j^m$ error suppressed (denoted by \red{R}) incorporate additional constraints~\cite{leung2018robust}, as described in Sec.~\ref{sec:preli}, to suppress detuning-induced errors, whereas the non-robust versions (denoted by NR) do not impose these constraints.

As derived in Eq.~\eqref{eq:infdmu_} of Appendix~\ref{app:detuning}, the first-order average infidelity under detuning error scales as
\begin{align}
1-\bar{F}\approx \frac{4}{5}\left[|\delta\alpha|^{2}+|\delta\Theta|^{2}\right],
\end{align}
where $|\delta\alpha|^{2}=\sum_{j,m}|\alpha_j^{m}(\tau)|^{2}$ is the total residual motional displacement.  In Fig.~\ref{fig:detuning}(c) and (d), we separate the error contribution from rotation angle and residual motional displacement. Other than some exceptional $\mu$, our methods provide significant reductions of both contributions.  This result verifies the robustness for both $|\delta\Theta|^2$ and $|\delta\alpha|^2$.

%Moreover, $|\delta\alpha|^{2}$ is significantly lower for S version for our methods, due to the same mechanism to~\cite{leung2018robust}.

% \red{\sout{The robustness against $|\delta\alpha|^2$ follows the similar mechanism to~\cite{leung2018robust}.}}
We now explain the mechanism of the robustness against $|\delta\Theta|^2$. As derived in Appendix~\ref{app:detuning}, the sensitivity of rotation angle is proportional to the off-diagonal part of $E$
\begin{align}\label{eq:ptpmu}
\frac{\partial\Theta}{\partial\mu}\sim \Gamma^{-1}\mathrm{Re}\left[E_{q,p}+E_{p,q}\right],
\end{align}
where the constant $\Gamma=\max_{m=1}^N\Gamma_m^{\uparrow}+\Gamma_m^{\downarrow}$ depends only on the heating error rate.
Besides, the cost function [see Eq.~\eqref{eq:estimation}] for our heating error minimization, $|E_{p,p}|+|E_{p,q}|$, is related to the diagonal part of $E$. Although not equal to each other, the diagonal and off-diagonal parts are closely related. As demonstrated in Theorem.~\ref{thm:thm1}, the cost function in Eq.~\eqref{eq:estimation} serves as an upper bound of the off-diagonal part, and hence the sensitivity in Eq.~\eqref{eq:ptpmu}. So minimizing heating error can at the same time reduce the sensitivity $\partial\Theta/\partial\mu$.

Furthermore, as shown in Fig.~\ref{fig:detuning}(d), our method yields a residual displacement $|\alpha|^2$ comparable to that of the min-Rabi method with the $\alpha$-robustness constraint enforced (yellow solid line), even without explicitly imposing this constraint (red dashed line). This behavior can be attributed to the sign-flip structure of the optimized waveform. As discussed in Ref.~\cite{hayes2012coherent}, waveforms resembling the Walsh-1 function can suppress residual motional displacement. This observation suggests that the $\alpha$-robustness constraints may be omitted without significantly degrading performance, thereby reducing the required number of segments $L$ and leading to a simpler waveform.

%To understand this improvement, we analyze both the rotation angle error, $|\delta\Theta|^2$, and total residual motional displacement, $|\delta\alpha|^{2}\equiv\sum_{j,m}|\alpha_j^{m}(\tau)|^{2}$. The result at  $\delta\mu=500\,\text{Hz}$ is shown in Fig.~\ref{fig:detuning}(c), (d).

%\red{To further elucidate the dominant sources of infidelity, we analyze the fluctuations in the accumulated rotation angle, $\delta\Theta$, and the total residual motional displacement, $|\delta\alpha|^{2}=\sum_{j,m}|\alpha_j^{m}(\tau)|^{2}$, at a fixed detuning error of $\delta\mu=500\,\text{Hz}$, as shown in Fig.~\ref{fig:detuning}(c) and (d). For both the robust min-Rabi method and our method, the infidelity scales as
%\begin{align}
%1-F\sim|\delta\Theta|^{2}\gg|\delta\alpha|^{2},
%\end{align}
% indicating that fluctuations in the accumulated rotation angle dominate the overall detuning error. In contrast, the constraints introduced in Ref.~\cite{leung2018robust} primarily ensure robustness of the residual displacements $\delta\alpha$. Since our cost function Eq.~\eqref{eq:bound} effectively suppresses the sensitivity
%\begin{align}
%\partial\Theta/\partial\mu\sim\mathrm{Re}\left[E_{q,p}+E_{p,q}\right],
%\end{align}
%our method yields substantially smaller rotation-angle fluctuations and, consequently, lower infidelity. See Appendix~\ref{app:detuning} for detailed derivation.}

\subsection{Compatibility with the Mitigation of Other Errors}
We have shown that our method is compatible with the detuning-error mitigation of Ref.~\cite{leung2018robust}, and provides additional robustness against detuning fluctuations $\delta\mu$. Beyond detuning errors, our framework is also readily compatible with a variety of other error-mitigation strategies. In appendix~\ref{app:combine}, we further give two examples. The first one is combining with the suppression of asymmetric errors~\cite{Zhang.25}. The second one is the combination with the amplitude smoothing technique in~\cite{cheng2023robust}. These results demonstrate the generality and flexibility of our heating error suppression approach.

\section{conclusions}
\label{sec:conclusion}
To conclude, we have proposed a general framework for mitigating heating error noise. Our framework works for different pulse basis, ion number, and system parameters. The framework is also compatible with the techniques for mitigation the error from the fluctuation of laser detuning.
% Our error suppression method is friendly experiment.
Our error suppression method is experimentally friendly.
For example, with the piecewise constant waveforms, the addition of few segments already gives a significant reduction of the heating errors.

Several improvements can be made based on our protocol. Firstly, one may optimize Eq.~\eqref{eq:bound} directly without the positive-extraction approximation. Although the global minimum of Eq.~\eqref{eq:bound} is hard to obtained in general, one may perform minimization with heuristic methods such as gradient descent~\cite{de.11} and reinforcement learning~\cite{zhang.19}. Secondly, the heating error bound Eq.~\eqref{eq:bound} only holds under the condition that the initial state is a product state of the qubit and phonon modes. While our method cannot ensure the final state is a product state, recooling is required after several gate operations. One may consider the general case where the initial state is a mixed state.
Thirdly, we have only considered quasi-static noise, and errors are assumed to be a constant during the gate operation in this work.

Future works may consider the mitigation of color noise, i.e. the error terms are time-dependent during the gate operation. The heating error suppression mechanism and its relation to other parameters, such as Rabi frequencies, also desire further studies. Finally, our method can be incorporated with entangling gate beyond Lamb-Dicke regime~\cite{orozco2025generally} to achieve both high fidelity and short gate time.

%Finally, due to the flexibility of our scheme, one may consider the combination of our technique to other techniques, such as holonomic gates~\cite{liang.14,ai.22} and compulsive pulse~\cite{Zhang.25}, to also improve the robustness to other type of noises.

\section*{acknowledgements}
We thank  Wenhao Zhang for helpful discussions. This work is supported by the National Natural Science Foundation of
China (No.~12405013 and No.~12361161602),
 NSAF (Grant
No.~U2330201), and Open
Fund of Key Laboratory of Atomic and Subatomic Structure
and Quantum Control (Ministry of Education).  The numerical simulation is supported by High-performance Computing
Platform of Peking University.

\begin{appendix}

\section{proof of theorems}
\label{app:proof}

\thmone*
\begin{proof}
Theorem~\ref{thm:thm1} utilizes Cauchy’s inequality and the arithmetic--geometric mean (AM-GM) inequality. For arbitrary phonon-mode trajectories $\alpha_j^m(t)$, we define a set of concatenated functions $F_j(x)$ that flatten the trajectory across all $N$ phonon modes into a single continuous function on $[0, N\tau]$:
\begin{equation}
    F_j(x)=\begin{cases}
        \sqrt{\Gamma_1^\uparrow+\Gamma_1^\downarrow}\alpha_j^1(x) \ &x\in[0,\tau) \\
        \sqrt{\Gamma_2^\uparrow+\Gamma_2^\downarrow}\alpha_j^2(x-\tau) \ &x\in[\tau,2\tau) \\
        \qquad\qquad \vdots \ & \vdots \\
        \sqrt{\Gamma_N^\uparrow+\Gamma_N^\downarrow}\alpha_j^N\left(x-(N-1)\tau\right) \ &x\in[(N-1)\tau, N\tau]
    \end{cases}.
\end{equation}
With this definition, the heating error bound in Eq.~\eqref{eq:bound} can be rewritten as
\begin{equation}
    \begin{aligned}
        E_{j_1,j_2}=&\sum_{m=1}^N\left(\Gamma_m^\uparrow+\Gamma_m^\downarrow\right)\int_0^\tau\mathrm{d}x\alpha_{j_1}^{m*}(\tau)\alpha_{j_2}^m(\tau) \\
        =&\int_0^{N\tau}\mathrm{d}xF_{j_1}^*(x)F_{j_2}(x).
    \end{aligned}
\end{equation}
Then, by applying Cauchy’s inequality and the AM-GM inequality, we obtain:
\begin{align}
    \left|E_{p,q}\right|+\left|E_{q,p}\right|=&2\left|\int_0^{N\tau}\mathrm{d}xF_{p}^*(x)F_{q}(x)\right| \\
    \leq&2\sqrt{\int_0^{N\tau}\mathrm{d}xF_{p}^*(x)F_{p}(x)} \notag\\
    &\times\sqrt{\int_0^{N\tau}\mathrm{d}xF_{q}^*(x)F_{q}(x)} \label{eq:ie1}\\
    =&2\sqrt{\left|E_{p.p}\right|\left|E_{q,q}\right|} \\
    \leq& \left|E_{p,p}\right|+\left|E_{q,q}\right|.\label{eq:ie2}
\end{align}
Here, inequality~\eqref{eq:ie1} follows from Cauchy--Schwarz, and inequality~\eqref{eq:ie2} from the AM-GM inequality. This completes the proof.
\end{proof}

\thmtwo*
\begin{proof}
To prove this theorem, we directly construct explicit forms of the matrices $\mathbf{S}_p,\mathbf{S}_q$ and $\tilde{\mathbf{M}}$.

Let $\mathcal{K}=\left\{\vec{c} \ | \ \mathbf{A}\vec{c}=\mathbf{A}^{\text{diff}}\vec{c}=0 \right\}$ be the simultaneous kernel space of $\mathbf{A}$ and $\mathbf{A}^{\text{diff}}$. Denote by $\mathbf{P}$ the orthogonal projection operator onto $\mathcal{K}$. The projected matrix $\mathbf{PMP}$ is symmetric and admits an eigendecomposition:
\begin{equation}
    \mathbf{PMP}=\sum_{k=1}^L \lambda_k\vec{v}_k\vec{v}_k^T,
\end{equation}
where $\lambda_k$ and $\vec{v}_k$ are the eigenvalues and eigenvectors of $\mathbf{PMP}$, respectively. Using these, we define a positive semidefinite matrix $\tilde{\mathbf{M}}$ as
\begin{equation}
    \tilde{\mathbf{M}}=\sum_{k=1}^L \left|\lambda_k\right|\vec{v}_k\vec{v}_k^T,
\end{equation}
which preserves the eigenvectors of $\mathbf{PMP}$ but replaces each eigenvalue with its absolute value. This operation can also be represented via a reflection matrix:
\begin{equation}
    \mathbf{F}=\sum_{k=1}^L\text{sign}(\lambda_k)\vec{v}_k\vec{v}_k^T,
\end{equation}
so that $\tilde{\mathbf{M}}=\mathbf{PMPF}$.Based on this, we define
\begin{equation}
    \mathbf{S}_p=\mathbf{P},\quad\mathbf{S}_q=\mathbf{PF}.
\end{equation}

The verification is straightforward once the matrices are constructed. For any vector $\vec{c}$ satisfying $\vec{c}^T \tilde{\mathbf{M}} \vec{c} = \Theta_{\text{targ}}$, let $\vec{c}_p = \mathbf{S}_p \vec{c}$ and $\vec{c}_q = \mathbf{S}_q \vec{c}$. It directly follows that:
\begin{enumerate}
    \item The constraint $\vec{c}_p^T \mathbf{M} \vec{c}_q = \Theta_{\text{targ}}$ is satisfied because
$\vec{c}_p^T \mathbf{M} \vec{c}_q = \vec{c}^T \mathbf{PMPF} \vec{c} = \vec{c}^T \tilde{\mathbf{M}} \vec{c} = \Theta_{\text{targ}}$;
    \item $\vec{c}_p$ and $\vec{c}_q$ lie in the kernel of both $\mathbf{A}$ and $\mathbf{A}^{\mathrm{diff}}$ because they are projections via $\mathbf{P}$.
\end{enumerate}

Thus, all three constraints are simultaneously satisfied, completing the proof.
\end{proof}

% After constructing, the proof of this theorem is obvious. Let a vector $\vec{c}$ satisfying $\vec{c}^T\mathbf{\tilde{M}}\vec{c}=\Theta_{\text{targ}}$, then define $\vec{c}_p=\mathbf{S}_p\vec{c}$ and $\vec{c}_q=\mathbf{S}_q\vec{c}$.
% First, the left-hand side of the constrain Eq.~\eqref{eq:sa} is
% \begin{equation}
%     \vec{c}_p^T\mathbf{M}\vec{c}_q=\vec{c}^T\mathbf{S}_p\mathbf{M}\mathbf{S}_q\vec{c}=\vec{c}^T\mathbf{PMPF}\vec{c}=\vec{c}\mathbf{\tilde{M}}\vec{c}=\Theta_{\text{targ}}
% \end{equation}
% since $\mathbf{\tilde{M}}=\mathbf{PMPF}$. Thus Eq.~\eqref{eq:sa} is satisfied.
% Next, we show that Eq.~\eqref{eq:sb} and Eq.~\eqref{eq:sc} are satisfied. Since $\mathbf{P}$ is the project operator onto the simultaneous kernel space $\mathrm{K}$ of $\mathbf{A}$ and $\mathbf{A}^{\text{diff}}$, it satisfies $\mathbf{AP}=\mathbf{A}^{\text{diff}}\mathbf{P}=0$. Hence,
% \begin{gather}
%     \mathbf{A}\vec{c}_p=\mathbf{AS}_p\vec{c}=\mathbf{AP}\vec{c}=0 \\
%     \mathbf{A}\vec{c}_q=\mathbf{AS}_q\vec{c}_q=\mathbf{APF}\vec{c}=0
% \end{gather}
% Thus, Eq.~\eqref{eq:sb} is satisfied. The constraint Eq.~\eqref{eq:sc} on $\mathbf{A}^{\text{diff}}$ are satisfied in a similar manner.

% To conclude, we reach the conclusion that whenever $\vec{c}$ satisfying $\vec{c}^T\mathbf{\tilde{M}}\vec{c}=\Theta_{\text{targ}}$, the three constrains of the optimization problem Eq.~\eqref{eq:m1} can be simultaneously satisfied.

\section{definitions and expressions}
\label{app:matexp}
In this appendix, we provide the explicit mathematical expressions underlying the gate construction, the heating error model, and the optimization framework introduced in the main text. These derivations clarify the assumptions and computational structures used throughout our method.

The time-dependent Hamiltonian generated by a set of individually addressed laser beams with Rabi frequencies $\Omega_j(t)$ and common detuning $\mu$ (applied to the $j$-th ion) is given by
\begin{equation}
    H(t) = \sum_{j=1}^N \sum_{m=1}^N \frac{i}{2} \eta_m b_j^m \Omega_j(t)
    \left( a_m^\dag e^{i\Delta_m t} - a_m e^{-i\Delta_m t} \right) \sigma_j,
    \label{eq:H}
\end{equation}
where $\eta_m$ is the Lamb-Dicke parameter, and $\vec{b}_j = (b_j^1, \dots, b_j^N)$ is the normalized eigenvector of the $m$-th motional mode for the $j$-th ion, both of which can be experimentally characterized~\cite{walther2012precision}. The offset $\Delta_m = \omega_m - \mu$ denotes the frequency difference between the $m$-th phonon mode and the laser detuning. Here, $a_m$ is the annihilation operator for the $m$-th mode, and $\sigma_j$ is the Pauli-$X$ operator acting on ion $j$ (for brevity, we omit the direction label "$X$").

Assuming only two ions $p$ and $q$ are illuminated (i.e., $\Omega_j(t) = 0$ for $j \neq p,q$), the resulting MS gate takes the form
\begin{equation}
    U(\tau) = \exp\left[\sum_{j=p,q} \phi_{j}(\tau) \sigma_j + i\,\Theta(\tau) \sigma_{p} \sigma_{q} \right],
    \label{eq:Utau}
\end{equation}
where the displacement operator $\phi_j(\tau)$ and the effective spin-spin coupling $\Theta(\tau)$ are defined as
\begin{align}
    \phi_j(\tau) =& \sum_{m=1}^N\left(\alpha_j^m(\tau) a_m^\dag - \alpha_j^{m*}(\tau) a_m \right), \\
    \alpha_j^m(\tau) =& \frac{1}{2} \eta_m b_j^m \int_0^\tau \Omega_j(t) e^{i\Delta_m t} \,\mathrm{d}t, \label{eq:alpha}\\
    \Theta(\tau) =& \frac{1}{4} \sum_{m=1}^N \eta_m^2 b_p^m b_q^m \int_0^\tau \mathrm{d}t_1 \int_0^{t_1} \mathrm{d}t_2\,  \notag \\
    &\left[ \Omega_p(t_1)\Omega_q(t_2) + \Omega_p(t_2)\Omega_q(t_1) \right] \sin \Delta_m(t_1 - t_2). \label{eq:Theta}
\end{align}
Here, $\alpha_j^m(\tau)$ denotes the trajectory of the $m$-th phonon mode in phase space induced by Rabi frequencies $\Omega_j(t)$, and $\Theta(\tau)$ characterizes the entangling strength between ions $p$ and $q$.

While the MS gate described above assumes unitary evolution under a time-dependent Hamiltonian, practical implementations are affected by decoherence processes such as motional heating. To model these effects more accurately, we adopt a master equation approach as follows:
\begin{equation}
    \frac{\partial\rho}{\partial t}=-i[H(t),\rho]+\sum_{m=1}^N\Gamma_m^\uparrow\mathcal{D}_m^\uparrow\left(\rho\right)+\Gamma_m^\downarrow\mathcal{D}_m^\downarrow\left(\rho\right),
    \label{eq:lindblad}
\end{equation}
where $\mathcal{D}_m^\downarrow(\rho) = a_m \rho a_m^\dagger - \frac{1}{2} \left\{ a_m^\dagger a_m, \rho \right\}$ and $\mathcal{D}_m^\uparrow(\rho) = a_m^\dagger \rho a_m - \frac{1}{2} \left\{ a_m a_m^\dagger, \rho \right\}$ are the standard Lindblad superoperators describing phonon relaxation and excitation, respectively, for the $m$-th motional mode. The coefficients $\Gamma_m^\uparrow$ and $\Gamma_m^\downarrow$ denote the corresponding excitation and relaxation rates.

Based on this dissipative dynamics, we define a cost function to estimate the infidelity due to heating and formulate an optimization problem to suppress it. To do so, we represent the control pulses in a chosen functional basis and express all relevant physical quantities in matrix form. For a given set of function basis $\vec{f}(t)=\left(f_1(t),f_2(t),\dots,f_L(t)\right)^T$, used to represent the Rabi frequencies $\Omega_j(t)$ via the expansion $\Omega_j(t) = \vec{c}\cdot\vec{f}(t)$, we now provide explicit expressions for the matrices $\mathbf{H}^{(j_1 j_2)}$, $\mathbf{M}$, $\mathbf{A}$, and $\mathbf{A}^{\text{diff}}$ introduced in Eq.~\eqref{eq:se}–\eqref{eq:sc} of the main text.

The heating cost matrix $\mathbf{H}^{(j_1 j_2)}$ appearing in the objective function is defined by
\begin{equation}
    \begin{aligned}
        \mathbf{H}^{(j_1 j_2)}_{l_1 l_2} = &\frac{1}{4} \sum_{m=1}^N \left( \Gamma_m^\uparrow + \Gamma_m^\downarrow \right) \eta_m^2 b_{j_1}^m b_{j_2}^m \\
        &\int_0^\tau \!\!\mathrm{d}t \int_0^t \!\! \mathrm{d}t_1 \int_0^t \!\! \mathrm{d}t_2 \, f_{l_1}(t_1) f_{l_2}(t_2) e^{i\Delta_m(t_2 - t_1)}.
    \end{aligned}
\end{equation}

The matrix $\mathbf{M}$ corresponds to the rotation angle $\Theta(\tau)$ and is given by
\begin{equation}
    \begin{aligned}
        \mathbf{M}_{l_1 l_2} = &\frac{1}{4} \sum_{m=1}^N \eta_m^2 b_p^m b_q^m \int_0^\tau \!\! \mathrm{d}t_1 \int_0^{t_1} \!\! \mathrm{d}t_2 \\
        &\left[ f_{l_1}(t_1) f_{l_2}(t_2) + f_{l_1}(t_2) f_{l_2}(t_1) \right] \sin\Delta_m(t_1 - t_2),
    \end{aligned}
\end{equation}
which is symmetric, i.e., $\mathbf{M}_{l_1 l_2} = \mathbf{M}_{l_2 l_1}$, but not necessarily positive semidefinite.

The constraint matrices $\mathbf{A}$ and $\mathbf{A}^{\text{diff}}$ arise from the frequency robustness conditions. Their entries are defined by
\begin{align}
    \mathbf{A}_{ml} = \int_0^\tau \!\! \mathrm{d}t \, f_l(t) e^{i\Delta_m t}, \quad
    \left(\mathbf{A}^{\text{diff}}\right)_{ml} = \int_0^\tau \!\! \mathrm{d}t \, t f_l(t) e^{i\Delta_m t}.
\end{align}

These matrices encode, respectively, the phonon displacement and its sensitivity to small mode frequency errors in each motional mode.

\section{derivation of heating error bound}
\label{app:errorest}
In Eq.\eqref{eq:bound} of the main text, we presented an upper-bound estimate for the heating-induced infidelity during the MS gate. In this section, we provide a detailed derivation of this bound, following the approach in Ref.\cite{he2024scaling}.

We consider the open-system dynamics governed by the Lindblad master equation, in contrast to the unitary evolution that yields the ideal gate. Specifically, the actual state $\rho(t)$ evolves as:
\begin{equation}
\frac{\partial\rho}{\partial t} = -i[H(t),\rho] + \sum_{m=1}^N \Gamma_m^\uparrow \mathcal{D}_m^\uparrow(\rho) + \Gamma_m^\downarrow \mathcal{D}_m^\downarrow(\rho),
\end{equation}
while the corresponding ideal state $\rho{\text{ideal}}(t)$ evolves unitarily under the Hamiltonian $H(t)$ as:
\begin{equation}
\frac{\partial\rho_{\text{ideal}}}{\partial t} = -i[H(t), \rho_{\text{ideal}}],
\end{equation}
with identical initial conditions:
\begin{equation}
\rho(0) = \rho_{\text{ideal}}(0)=\rho_0.
\end{equation}
Here, we adopt the same definition of the dissipators and noise rates as Eq.~\eqref{eq:lindblad} introduced in the previous section. Specifically, the excitation and relaxation processes $\mathcal{D}_m^\uparrow(\cdot)$ and $\mathcal{D}_m^\downarrow(\cdot)$, as well as their associated rates $\Gamma_m^\uparrow$ and $\Gamma_m^\downarrow$, are defined therein.

We define the heating error as the infidelity between the reduced states of $\rho(\tau)$ and $\rho_{\text{ideal}}(\tau)$ at the end of the gate duration $\tau$, after tracing out the phonon degrees of freedom:
\begin{equation}
    \epsilon_{\text{heating}} := 1 - F\left(\tr_{\text{ph}}\rho(\tau), \tr_{\text{ph}}\rho_{\text{ideal}}(\tau)\right),
    \label{eq:ptrinfid}
\end{equation}
where $F(\rho_1, \rho_2)$ denotes the fidelity between two density matrices. We assume throughout that the Rabi frequencies has been optimized to implement the desired MS gate in the absence of noise, and hence, the remaining infidelity is dominantly caused by motional heating.

To estimate the heating error, we first move to the interaction picture with respect to the time-dependent Hamiltonian $H(t)$, defined in Eq.~\eqref{eq:H}. The density matrices transform as
\begin{align}
    \rho(t) &\rightarrow \tilde{\rho}(t) = U^\dag(t) \rho(t) U(t), \\
    \rho_{\text{ideal}}(t) &\rightarrow \tilde{\rho}_{\text{ideal}}(t) = U^\dag(t) \rho_{\text{ideal}}(t) U(t) = \rho_0,
\end{align}
where $U(t)$ is the unitary evolution operator generated by $H(t)$. In this frame, the ideal state becomes time-independent, while the full dynamics of $\tilde{\rho}(t)$ is governed solely by the dissipative processes.

The Lindblad master equation in the interaction picture becomes:
\begin{equation}
    \frac{\partial\tilde{\rho}(t)}{\partial t}
    =\tilde{\mathcal{L}}(\rho(t))= \sum_{m=1}^N \Gamma_m^\uparrow \tilde{\mathcal{D}}_m^\uparrow\left(\tilde{\rho}(t)\right)
    +\Gamma_m^\downarrow\tilde{\mathcal{D}}_m^\downarrow\left(\tilde{\rho}(t)\right),
\end{equation}
where the interaction-picture dissipators are defined as
\begin{align}
    \tilde{\mathcal{D}}_m^\downarrow(\rho) &= \tilde{a}_m(t)\rho \tilde{a}_m^\dag(t) - \frac{1}{2}\left\{\tilde{a}_m^\dag(t) \tilde{a}_m(t), \rho \right\}, \\
    \tilde{\mathcal{D}}_m^\uparrow(\rho) &= \tilde{a}_m^\dag(t)\rho \tilde{a}_m(t) - \frac{1}{2}\left\{\tilde{a}_m(t) \tilde{a}_m^\dag(t), \rho \right\},
\end{align}
and the transformed phonon annihilation operator is
\begin{equation}
\tilde{a}_m(t) = U^\dag(t) a_m U(t) = a_m + \sum_{j=1}^N \alpha_j^m(t) \sigma_j
\equiv a_m + S_m(t).
\end{equation}

Assuming that the noise is weak, i.e. $\|\Lambda\|=\left\|\int_0^\tau\mathrm{d}t\tilde{\mathcal{L}}(\rho_0)\right\|\ll1$, the evolved density matrix can be expanded as:
\begin{equation}
    \tilde{\rho}(\tau)=\tilde{\rho}_{\text{ideal}}(\tau)+\Lambda+O(\|\Lambda\|^2).
\end{equation}
The heating error, defined as Eq.~\eqref{eq:ptrinfid}, can be estimated as
\begin{align}
    \epsilon_{\text{heating}}&\leq\frac{1}{2}\left\|\tr_{\text{ph}}\left(\rho(\tau)-\rho_{\text{ideal}}(\tau)\right)\right\| \\
    &=\frac{1}{2}\left\|\tr_{\text{ph}}\left(U(\tau)\tilde{\rho}(\tau)U(\tau)^\dag-U(\tau)\tilde{\rho}_{\text{ideal}}(\tau)U(\tau)^\dag\right)\right\| \\
    &\lesssim\frac{1}{2}\left\|\tr_{\text{ph}}\left(U(\tau)\Lambda U(\tau)^\dag\right)\right\|.
\end{align}
Since $H(t)$ is assumed to implement the target gate, $U(\tau) \approx U_{\text{ideal}}(\tau) = \exp(i\Theta_{\text{targ}} \sigma_p \sigma_q)$ acts trivially on the phonon Hilbert space. Therefore, $U(\tau)$ commutes with the partial trace and the error simplifies to
\begin{align}
    \epsilon_{\text{heating}}\lesssim&\frac{1}{2}\left\|U(\tau)\tr_{\text{ph}}\Lambda U^\dag(\tau)\right\| \\
    =&\frac{1}{2}\left\|\tr_{\text{ph}}\Lambda\right\| \\
    =&\frac{1}{2}\left\|\sum_{m=1}^N\Gamma_m^\uparrow\int_0^\tau\mathrm{d}t\tr_{\text{ph}}\left(\tilde{\mathcal{D}}_m^\uparrow(\rho_0)\right)\right. \notag\\
    &\left. +\sum_{m=1}^N\Gamma_m^\downarrow\int_0^\tau\mathrm{d}t\tr_{\text{ph}}\left(\tilde{\mathcal{D}}_m^\downarrow(\rho_0)\right)\right\|.
\end{align}
Let us focus on the excitation term (the analysis for the relaxation term is analogous). Using the decomposition $\tilde{a}_m(t) = a_m + S_m(t)$, the integrand can be expanded into four terms:
\begin{align}
    &\sum_{m=1}^N\Gamma_m^\uparrow\int_0^\tau\!\!\mathrm{d}t\tr_{\text{ph}}\tilde{\mathcal{D}}_m^\uparrow(\rho_0) \\
    =&\sum_{m=1}^N\Gamma_m^\uparrow\int_0^\tau\!\!\mathrm{d}t\tr_{\text{ph}}\left(a_m^\dag\rho_0a_m-\frac{1}{2}\{a_ma_m^\dag,\rho_0\}\right) \label{eq:aaterm}\\
    +&\sum_{m=1}^N\Gamma_m^\uparrow\int_0^\tau\!\!\mathrm{d}t\tr_{\text{ph}}\left(a_m^\dag\rho_0S_m(t)-\frac{1}{2}\{S_m(t)a_m^\dag,\rho_0\}\right) \label{eq:asterm} \\
    +&\sum_{m=1}^N\Gamma_m^\uparrow\int_0^\tau\!\!\mathrm{d}t\tr_{\text{ph}}\left(S_m^\dag(t)\rho_0a_m-\frac{1}{2}\{a_mS_m^\dag(t),\rho_0\}\right) \label{eq:saterm} \\
    +&\sum_{m=1}^N\Gamma_m^\uparrow\int_0^\tau\!\!\mathrm{d}t\tr_{\text{ph}}\left(S_m^\dag(t)\rho_0S_m(t)-\frac{1}{2}\{S_m(t)S_m^\dag(t),\rho_0\}\right). \label{eq:ssterm}
\end{align}
Under the assumption that the initial state $\rho_0$ is a product state of the spin and phonon degrees of freedom, the mixed terms Eq.~\eqref{eq:asterm} and Eq.~\eqref{eq:saterm} vanish upon taking the partial trace. Additionally, the phonon-only term in Eq.~\eqref{eq:aaterm} yields zero due to bosonic trace properties. Hence, only the spin-dependent term in Eq.~\eqref{eq:ssterm} contributes to the heating error.

\begin{figure}[t]
    \centering
    \includegraphics[width=\columnwidth]{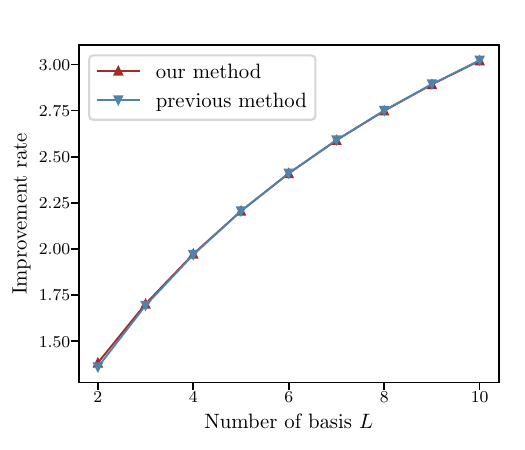}
    % \caption{Improve rate of our method and previous method \red{using polychromatic basis} with 2 ions, excitation and relaxation rate $\Gamma^\uparrow=\Gamma^\downarrow=500\text{ phonon}/s$, basic detuning $\Delta=2\pi\times0.05\,\text{MHz}$, rotation angle $\Theta_{\text{ideal}}=\pi/4$, gate duration $\tau=\frac{2\pi}{\Delta}$. \red{Blue line: previous method given in Eq.~\eqref{eq:pre_method} by~\cite{Haddadfarshi_2016}; red line: our new method given in Eq.~\eqref{eq:our_method}.}}
    \caption{Improvement rate of our method compared with a previous approach, both formulated in the Fourier basis, for a two-ion system within a simplified single-mode model. The excitation and relaxation rates are $\Gamma^{\uparrow}=\Gamma^{\downarrow}=500\,\text{phonons/s}$, the base detuning is $\Delta = 2\pi \times 0.05\,\text{MHz}$, and the gate duration is $\tau = 2\pi/\Delta$. The blue curve corresponds to the previous method proposed in Ref.~\cite{Haddadfarshi_2016}, as defined in Eq.~\eqref{eq:pre_method}, while the red curve represents our method given in Eq.~\eqref{eq:our_method}.}
    \label{fig:compare}
\end{figure}
% \begin{figure}[t]
%     \centering
%     \includegraphics[width=\columnwidth]{fig/polychrom.pdf}
%     \caption{\red{Heating-error estimation for our method using Fourier and piecewise-constant bases in the full multi-mode model with $N=4$ ions. The gate duration is $\tau = 150,\mu\text{s}$, and the excitation and relaxation rates are $\Gamma_m^{\uparrow,\downarrow} = 100,\text{phonons/s}$. The blue curve corresponds to the Fourier basis, while the red curve corresponds to the piecewise-constant basis used throughout the main text.}}
%     \label{fig:polychrom_compare}
% \end{figure}
% \begin{figure}[t]
%     \centering
%     \includegraphics[width=\columnwidth]{fig/polychrom_bad.pdf}
%     \caption{\red{}}
%     \label{fig:polychrom_compare_good}
% \end{figure}
% \begin{figure}[t]
%     \centering
%     \includegraphics[width=\columnwidth]{fig/polychrom_good.pdf}
%     \caption{\red{}}
%     \label{fig:polychrom_compare_bad}
% \end{figure}
\begin{figure}[t]
    \centering
    \includegraphics[width=\linewidth]{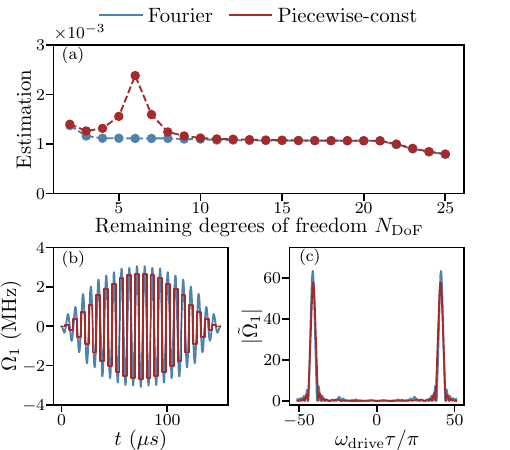}
    \caption{Comparison of Fourier~\eqref{eq:fourier_basis} and piecewise-constant basis for a full multi-mode model with $N=4$ ions. In both basis, the Rabi frequency $\Omega(t)$ is optimized using our method. The gate duration is $\tau = 150,\mu\text{s}$, the laser detuning is $\mu = \omega_{\mathrm{COM}} + 2\pi \times 0.002\,\text{MHz}$, and the heating rates are $\Gamma_m^{\uparrow,\downarrow} = 100\,\text{phonons/s}$.
    Panels (a) shows the estimated heating error as a function of the remaining degrees of freedom $N_{\mathrm{Dof}}$. Panels (b) and (c) illustrate a representative case with $N_{\mathrm{DoF}}=25$: (b) shows the optimized Rabi frequency of the first ion $\Omega_1(t)$, while (c) shows the corresponding spectra $\tilde{\Omega}_1(\omega_{\mathrm{drive}})=\frac{1}{\sqrt{2\pi}}\int_0^\tau \Omega_1(t)e^{-i\omega_{\mathrm{drive}} t}\,dt$.
    In all subfigures, blue curves correspond to the Fourier basis, and red curves correspond to the piecewise-constant basis used throughout the main text.}
    \label{fig:polychrom_compare}
\end{figure}

Define the inner qubit state as $\rho^{\text{in}}_0 = \tr_{\text{ph}} \rho_0$. We now estimate the excitation contribution to the heating error by analyzing the spin-dependent part in Eq.~\eqref{eq:ssterm}. Substituting the expression of $S_m(t)$ and using linearity of the trace, we obtain
\begin{align}
    &\left\|\sum_{m=1}^N\Gamma_m^\uparrow\int_0^\tau\mathrm{d}t\tr_{\text{ph}}\tilde{\mathcal{D}}_m^\uparrow(\rho_0)\right\| \\
    =&\left\|\sum_{m=1}^N\Gamma_m^\uparrow\!\!\int_0^\tau\!\!\!\!\mathrm{d}t\left(S_m^\dag(t)\rho^{\text{in}}_0S_m(t)-\frac{1}{2}\{S_m(t)S_m^\dag(t),\rho^{\text{in}}_0\}\right)\right\| \\
    \leq&\sum_{j_1,j_2=1}^N\left|\sum_{m=1}^N\Gamma_m^\uparrow\int_0^\tau\mathrm{d}t\alpha_{j_1}^{m*}(t)\alpha_{j_2}^m(t)\right| \notag\\
    &\phantom{\sum_{j_1,j_2=1}^N}\times\left(\left\|\sigma_{j_1}\rho^{\text{in}}_0\sigma_{j_2}\right\|+\frac{1}{2}\left\|\sigma_{j_2}\sigma_{j_1},\rho^{\text{in}}_0\right\|\right) \\
    \leq&2\sum_{j_1,j_2=1}^N\left|\sum_{m=1}^N\Gamma_m^\uparrow\int_0^\tau\mathrm{d}t\alpha_{j_1}^{m*}(t)\alpha_{j_2}^m(t)\right|.
\end{align}
An identical argument applies to the relaxation term, yielding
\begin{equation}
    \begin{aligned}
        &\left\|\sum_{m=1}^N\Gamma_m^\downarrow\int_0^\tau\mathrm{d}t\tr_{\text{ph}}\left(\tilde{\mathcal{D}}_m^\downarrow(\rho_0)\right)\right\| \\
        \leq&2\sum_{j_1,j_2=1}^N\left|\sum_{m=1}^N\Gamma_m^\downarrow\int_0^\tau\mathrm{d}t\alpha_{j_1}^{m*}(t)\alpha_{j_2}^m(t)\right|.
    \end{aligned}
\end{equation}
Adding both contributions, we obtain the final upper bound:
\begin{equation}
    \begin{aligned}
        &1-F(\tr_{\text{ph}}\rho(\tau),\tr_{\text{ph}}\rho_{\text{ideal}}(\tau)) \\
        \leq&\sum_{j_1,j_2=1}^N\left|\sum_{m=1}^N\left(\Gamma_m^\uparrow+\Gamma_m^\downarrow\right)\int_0^\tau\mathrm{d}t\alpha_{j_1}^{m*}(t)\alpha_{j_2}^m(t)\right|.
    \end{aligned}
\end{equation}

\section{comparison with polychromatic method}
\label{app:premethod}
Haddadfarshi et al.~\cite{Haddadfarshi_2016} proposed a method for minimizing the heating error under a simplified model, where only the center-of-mass (COM) phonon mode is considered, and the laser illumination is assumed to be identical across all ions. In this appendix, we show that our method reduces to their approach under the same model, thereby establishing a direct theoretical connection.

Under the simplified single-mode model, the Hamiltonian and dissipation term reduce to:
\begin{gather}
    H(t)=\left(\Upsilon(t)a^\dag+\Upsilon^*(t)a\right)\sum_{j=1}^N\sigma_j \\
    \mathcal{L}\left(\rho(t)\right)=\Gamma\left(\mathcal{D}^\uparrow\left(\rho(t)\right)+\mathcal{D}^\downarrow\left(\rho(t)\right)\right)
\end{gather}
where $\Upsilon(t)$ denotes the common Rabi frequencies applied to all ions, and $\Gamma$ is the heating rate of the single (usually COM) motional mode. $\mathcal{D}^\uparrow\left(\rho(t)\right)=a^\dag\rho(t)a-\frac{1}{2}\{aa^\dag,\rho(t)\}$ and $\mathcal{D}^\downarrow\left(\rho(t)\right)=a\rho(t)a^\dag-\frac{1}{2}\{a^\dag a,\rho(t)\}$ are the excitation and relaxation processes, respectively.
In our main text, we use a monochromatic laser, corresponding to $\Upsilon(t) = \frac{i}{2} \eta b \Omega(t) e^{i\Delta t}$. In contrast, Ref.~\cite{Haddadfarshi_2016} considered a polychromatic driving field composed of multiple detuning components:
\begin{equation}
    \Upsilon(t)=\sum_{l=1}^L c_l\Delta e^{il\Delta t}
\end{equation}
where $c_l$ denotes the coefficient of the $l$-th frequency tone with detuning $l\Delta$. This choice implicitly defines a Fourier basis $f_l(t) = \Delta e^{i l \Delta t}$, such that the driving profile $\Upsilon(t)$ is parameterized by a coefficient vector $\vec{c} = (c_1, \dots, c_L)$.

Under this model, the optimization constraints become:
\begin{gather}
    \alpha(\tau)=\sum_{l=1}^L\frac{c_l}{l}\left(1-e^{il\Delta\tau}\right)=0 \\
    \Theta(\tau)=2\sum_{k,l=1}^Lc_kc_l\left(\frac{\sin(k-l)\Delta\tau}{l(k-l)}-\frac{\sin k\Delta\tau}{kl}\right)=\Theta_{\text{targ}}
\end{gather}
Here, $\alpha(\tau)$ is the residual displacement of the phonon mode and $\Theta(\tau)$ is the spin–spin interaction angle. We omit robustness conditions such as $\partial_\Delta \alpha(\tau) = 0$, since they are also neglected in the comparison method. By choosing the gate duration as $\tau = \tfrac{2\pi}{\Delta}$, the displacement constraint becomes $\alpha(\tau) = 0$ automatically. Under this choice, the expression for $\Theta(\tau)$ simplifies to:
\begin{equation}
\Theta(\tau) = 4\pi \sum_{l=1}^L \frac{c_l^2}{l} = \Theta_{\text{targ}}.
\end{equation}
Using the heating error bound from Eq.~\eqref{eq:bound}, the heating error can be approximated by:
\begin{gather}
    E=\Gamma\frac{2\pi}{\Delta}\left(\sum_{l=1}^L\frac{c_l^2}{l^2}+\left(\sum_{l=1}^L\frac{c_l}{l}\right)^2\right)
\end{gather}

Thus, when our method is applied under the simpler model considered in Ref.~\cite{Haddadfarshi_2016}, i.e., only the center-of-mass mode is included and identical Rabi frequency is applied to all ions, the optimization problem reduces to:
\begin{equation}
    \begin{aligned}
        \minimize_{c_l}\quad &\sum_{l=1}^L\frac{c_l^2}{l^2}+\left(\sum_{l=1}^L\frac{c_l}{l}\right)^2 \\
    \text{subject to}\quad &4\pi\sum_{l=1}^L\frac{c_l^2}{l}=\Theta_{\text{ideal}}
    \end{aligned}
    \label{eq:our_method}
\end{equation}
This is closely related to the formulation in Ref.~\cite{Haddadfarshi_2016}, which instead uses:
\begin{equation}
    \begin{aligned}
        \minimize_{c_l}\quad&\sum_{l=1}^L\frac{c_l^2}{l^2} \\
    \text{subject to}\quad&4\pi\sum_{l=1}^L\frac{c_l^2}{l}=\Theta_{\text{ideal}} \\
    &\sum_{l=1}^L\frac{c_l}{l}=0
    \end{aligned}
    \label{eq:pre_method}
\end{equation}
Here, the extra constraint $\sum_l \tfrac{c_l}{l} = 0$ was imposed explicitly to suppress several terms in the error expression, whereas in our formulation this term is directly minimized as part of the objective.

Despite these differences in formulation, both methods yield almost identical improvements in heating error, as confirmed in Fig.~\ref{fig:compare}. However, we note that the waveform produced by our method often achieves slightly better performance when the number of components $L$ is small.

% \red{Using our framework, the polychromatic approach developed for simplified models can be formally generalized to the full multi-mode setting. In Fig.~\ref{fig:polychrom_compare}, we compare the estimated heating errors obtained using Fourier and piecewise-constant bases. Although the Fourier basis yields a lower theoretical heating error, this advantage is primarily formal. In a system with $M>1$ phonon modes, at least $L_{\min}=2M+1$ Fourier components are required, implying the need for a large number of simultaneously applied laser beams to realize a heating-robust MS gate. Such requirements pose significant experimental challenges, making the Fourier-basis implementation impractical compared with piecewise-constant waveforms.}
\begin{figure*}
    \centering
    \includegraphics*[width=0.95\linewidth]{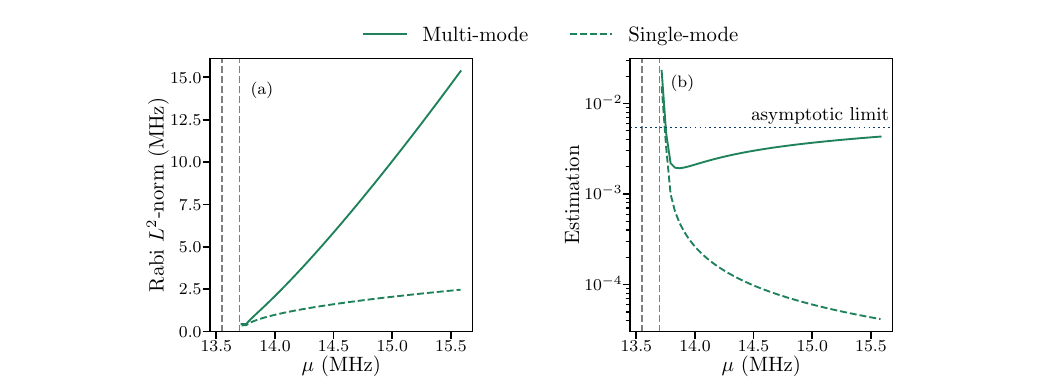}
    \caption{Rabi $L^2$-norm and heating error estimation s functions of the laser detuning $mu$ for the multi-mode and single-mode models. The solid and dashed green curves correspond to the multimode and single-mode cases, respectively. The two vertical dashed gray lines indicate the frequencies of the two highest-frequency motional modes, namely the center-of-mass (COM) mode and the breathing mode. The dotted blue line in (b) denotes the large-detuning asymptotic limit of the multimode model predicted by Eq.~\eqref{eq:adlimit}.}
    \label{fig:adlimit}
\end{figure*}
Using our framework, the polychromatic approach originally developed for simplified models can be formally extended to the full multi-mode setting. Instead of adopting the piecewise-constant basis in Eq.~\eqref{eq:piecewisebasis}, we consider the following Fourier basis functions:
\begin{equation}
\begin{aligned}
%     f_l^{s}(t)=&\sin\!\left[(l+1)\pi\frac{t}{\tau}\right], \\
% f_l^{c}(t)=&\cos\!\left[\left(l+\tfrac{1}{2}\right)\pi\frac{t}{\tau}\right],
    f_l^{\mathrm{Fourier}}=\cos\left[(l-1)\pi\frac{t}{\tau}\right]
\end{aligned}
\end{equation}
and expand the Rabi frequency as
\begin{equation}
\Omega(t)
=\sum_{l=1}^{L} c_l\, f_l^{\mathrm{Fourier}}(t).
\label{eq:fourier_basis}
\end{equation}
To obtain more experimentally realistic waveforms, we impose two additional constraints such that the pulse amplitude vanishes at both the beginning and the end of the gate:
\begin{equation}
    \sum_{l=1}^L c_l=\sum_{l=1}^L (-1)^{l-1}c_l=0.
    \label{eq:fourier_constrain}
\end{equation}
Since the Fourier basis introduces two additional linear constraints, comparing the two bases using the same number of basis functions $L$ would not correspond to the same optimization complexity. Therefore, we instead compare them at the same number of remaining optimization degrees of freedom, denoted by $N_{\mathrm{DoF}}$. Specifically,
\begin{equation}
    N_{\mathrm{DoF}}=\begin{cases}
        L-4N, & \text{piecewise-constant basis},\\
        L-4N-2, & \text{Fourier basis},
    \end{cases}
\end{equation}
where the additional reduction by two in the Fourier basis comes from the endpoint constraints in Eq.~\eqref{eq:fourier_constrain}. The comparison results are shown in Fig.~\ref{fig:polychrom_compare}.
% We compare this alternative basis with the piecewise-constant basis in Fig.~\ref{fig:polychrom_compare}.
% Two different laser detunings $\mu$ are considered: the left and right parts correspond to the two detuning values, respectively. For each case, we show the heating-error estimation as well as the corresponding Rabi frequencies in both the time and frequency domains.

Figure~\ref{fig:polychrom_compare} shows that, after optimization, the two bases converge to nearly identical waveforms and achieve almost the same level of heating-error suppression. The main difference is that the waveform obtained with the Fourier basis exhibits a slightly higher peak Rabi frequency. This is because the Fourier basis produces smoother waveforms, which spend less time near their peak amplitude than piecewise-constant waveforms. Consequently, to achieve the same entangling phase—which is approximately determined by the $L^2$-norm of the waveform—a slightly larger peak Rabi frequency is required. On the other hand, panel (a) shows that the optimization using the Fourier basis is more stable, with a weaker dependence on the number of optimization degrees of freedom.

These results demonstrate that our optimization framework is not tied to a particular choice of basis functions. Instead, it can be naturally applied to both piecewise-constant and smooth parameterizations while achieving comparable heating-error suppression. Therefore, different bases can be chosen according to experimental convenience—for example, depending on whether piecewise-constant or smoothly varying waveforms are easier to implement—without sacrificing the performance of the optimization.

% The Rabi frequencies optimized using the Fourier basis are relatively smaller in amplitude, and their spectra are more strongly concentrated near zero frequency. However, the optimization performance is highly sensitive to the laser detuning: a detuning shift of $2\pi\times 0.0025\,\text{MHz}$ increases the heating error from $3\times10^{-4}$ to $3.5\times10^{-3}$ and leads to substantial changes in the optimal waveforms.

% In contrast, for the piecewise-constant basis, the heating error stabilizes at approximately $7\times10^{-4}$, and the optimized waveforms remain qualitatively similar across different detunings, indicating significantly enhanced robustness. Although the piecewise-constant waveforms exhibit rapid oscillations, these correspond only to fast phase switching between $0$ and $\pi$. Such phase switching is typically treated as ideal in experiments, making these waveforms equally straightforward to implement in practice.

\section{comparison with detuning-rescaling method}
\label{app:detuning_rescaling}
Another well-established approach for suppressing heating errors is based on the detuning-rescaling~\cite{molmer1999multiparticle,sorensen1999quantum}, which increases the laser detuning $\mu$ together with the Rabi frequency $\Omega$.

% \begin{figure*}
%     \centering
%     \includegraphics*[width=0.95\linewidth]{fig/infid_rabi_N4.pdf}
%     \caption{\red{Heating error estimation and Rabi $L^2$-norm for different pairs of ions in a four-ion chain ($N=4$) The first column shows the heating-error estimation as a function of the Rabi $L^2$-norm. The second and third columns show the heating-error estimation and the Rabi $L^2$-norm, respectively, as functions of the laser detuning $\mu$. In all subfigures, the green curves correspond to the detuning-rescaling method, while the red curves correspond to our method. The two vertical dashed gray lines indicate the frequencies of the two highest-frequency motional modes, namely the center-of-mass (COM) mode and the breathing mode. The dotted blue line denotes the large-detuning asymptotic limit of the multi-mode model predicted by Eq.~\eqref{eq:adlimit}.}}
%     \label{fig:infid_rabi}
% \end{figure*}

Specifically, consider the pulse
\begin{equation}
    \Omega_p(t)=\Omega_q(t)=\Omega \left(1-\cos\frac{2\pi t}{\tau}\right).
\end{equation}
The original proposal considered only single mode, so we begin with assuming  that only the COM mode is involved, i.e., the summation over $m$ in Eqs.~\eqref{eq:Theta} and~\eqref{eq:bound} contains a single term. Assuming that the pulse satisfies the adiabatic condition $2\pi/\tau\ll|\Delta_{\mathrm{COM}}|$, the accumulated entangling phase and the heating error can be approximated by
\begin{align}
    \Theta_{\text{single}}&\approx\frac{3}{4}\eta^2_{\text{COM}}b^{\text{COM}}_pb^{\text{COM}}_q\frac{\Omega^2\tau}{\Delta_{\text{COM}}}, \\
    E_{\text{single}}&\approx\frac{3}{8}\eta_{\text{COM}}^2\left[(b^{\text{COM}}_p)^2+(b^{\text{COM}}_q)^2\right]\Gamma\frac{\Omega^2\tau}{\Delta_{\text{COM}}^2}.
\end{align}
Therefore, increasing $\Delta{\mathrm{COM}}$ by a factor of $K$ while increasing $\Omega$ by $\sqrt{K}$ suppresses the heating error by a factor of $K$ while keeping the entangling phase unchanged.

% Increasing Rabi frequency, however, may introduce additional experimental errors, such as photon scattering and laser control errors. This naturally motivates comparing different methods using the normalized quantity $E/\Omega^2$, rather than $E$ alone.
In this appendix,
% we address two questions. First,
we show that the detuning-rescaling strategy does not directly extend to multimode systems.
% Second, we demonstrate that our method still outperforms conventional approaches even when compared using the normalized metric $E/\Omega^2$.

\subsection{Error convergence in multi-mode case}

We now consider the large-detuning limit for a multimode system. Since $|\omega_{m}-\omega_{\text{COM}}|\ll|\Delta_{\text{COM}}|=|\omega_{\text{COM}}-\mu|$, the detuning of each mode can be expanded as
\begin{equation}
    \Delta_m=\Delta_{\mathrm{COM}}-(\omega_{\mathrm{COM}}-\omega_m),
\end{equation}
and, to leading order,
\begin{equation}
    \frac{1}{\Delta_m}=\frac{1}{\Delta_{\mathrm{COM}}}+\frac{\omega_{\mathrm{COM}}-\omega_m}{\Delta_{\mathrm{COM}}^2}+O(\Delta_{\mathrm{COM}}^{-3}).
\end{equation}
In this subsection, we assume that the Lamb--Dicke parameters are fixed $\eta_m=\eta$. The actual maximal fluctuation of $\eta_m$ is only on the order of $0.01$, and numerical verification confirms that this variation is negligible.
%Moreover, the Lamb--Dicke parameters are approximately mode independent, $\eta_m\approx\eta$.
Substituting the leading-order term into Eq.~\eqref{eq:Theta}, the entangling phase becomes
\begin{equation}
    \Theta_{\text{multi}}=\frac{3\eta^2\Omega^2\tau}{4}\sum_{m=1}^N \frac{b^m_pb^m_q}{\Delta_m}=\frac{3\eta^2\Omega^2\tau}{4\Delta_{\text{COM}}}\sum_{m=1}^N b^m_pb^m_q=0
\end{equation}
where we have used the orthogonality of the normal-mode vectors, $\sum_m b_p^mb_q^m=\delta_{pq}=0$ for $p\neq q$. Therefore, the leading-order contribution vanishes, and the next-order term must be retained:
\begin{equation}
    \Theta_{\text{multi}}=\frac{3\eta^2\Omega^2\tau}{4\Delta^2_{\text{COM}}}\sum_{m=1}^N b^m_pb^m_q\left(\omega_{\text{COM}}-\omega_m\right).
\end{equation}
Meanwhile, the leading-order heating error does not vanishes
\begin{equation}
    E_{\text{multi}}=\frac{3\eta^2\Gamma\Omega^2\tau}{8\Delta^2_{\text{COM}}}\sum_{m=1}^N\left[(b_p^m)^2+(b_q^m)^2\right].
\end{equation}
Therefore, in the large-detuning limit, both the entangling phase and the heating error scale as $O(\Omega^2/\Delta_{\mathrm{COM}}^2)$. Consequently, unlike the single-mode case, increasing the detuning while keeping the entangling phase fixed no longer suppresses the heating error. Instead, the heating error approaches the asymptotic limit
\begin{equation}
    E_{\text{multi}}\to\frac{\Gamma\sum_{m=1}^N\left[(b_p^m)^2+(b_q^m)^2\right]}{2\sum_{m=1}^N b^m_pb^m_q\left(\omega_{\text{COM}}-\omega_m\right)}\Theta_{\text{multi}},
    \label{eq:adlimit}
\end{equation}
which shows that, unlike the single-mode case, the heating error no longer decreases indefinitely with increasing detuning, but instead converges to a finite value proportional to the target entangling phase.

\subsection{Numerical comparison}

In Fig.~\ref{fig:adlimit}, we compare the detuning-rescaling method for single-mode and multi-mode models. As expected, the required Rabi frequency scales as  $\sqrt{\Delta_{\mathrm{COM}}}$ in the single-mode case, whereas it scales linearly with $\Delta_{\mathrm{COM}}=\omega_{\mathrm{COM}}-\mu$ in the multimode case in order to maintain a fixed entangling phase. Consequently, as the detuning increases, the heating-error estimation for the multimode case converges to the large-detuning asymptotic limit, and the numerical results are in excellent agreement with the theoretical prediction of Eq.~\eqref{eq:adlimit}.

% For multimode systems, we further compare our method with the detuning-rescaling method by evaluating the heating error under the same Rabi $L^2$-norm. The first column of Fig.~\ref{fig:infid_rabi} shows the heating-error estimation as a function of the Rabi $L^2$-norm for several pairs of ions. For non-nearest-neighbor ion pairs, our method consistently achieves lower heating errors than the detuning-rescaling method. For nearest-neighbor pairs, however, the performance of the two methods is comparable. As the Rabi frequency increases, the results gradually converge. This is expected because, in the large-detuning limit, the detailed temporal structure of the waveform becomes less important, and only the overall scaling with the Rabi $L^2$-norm remains.

% Nevertheless, our method retains a significant practical advantage. As shown above, the detuning-rescaling strategy cannot suppress the heating error arbitrarily in multimode systems. Numerically, we find that for non-nearest-neighbor ion pairs this heating error remains above $10^{-4}$. By contrast, our method is able to reduce the heating error below $10^{-4}$ while requiring only a moderate Rabi frequency, a performance that cannot be achieved by the detuning-rescaling strategy.

\section{Infidelities and $L_2$-norms of Rabi frequencies}
\label{app:infid&l2norm}
\begin{figure*}[t]
    \centering
    \includegraphics[width=\linewidth]{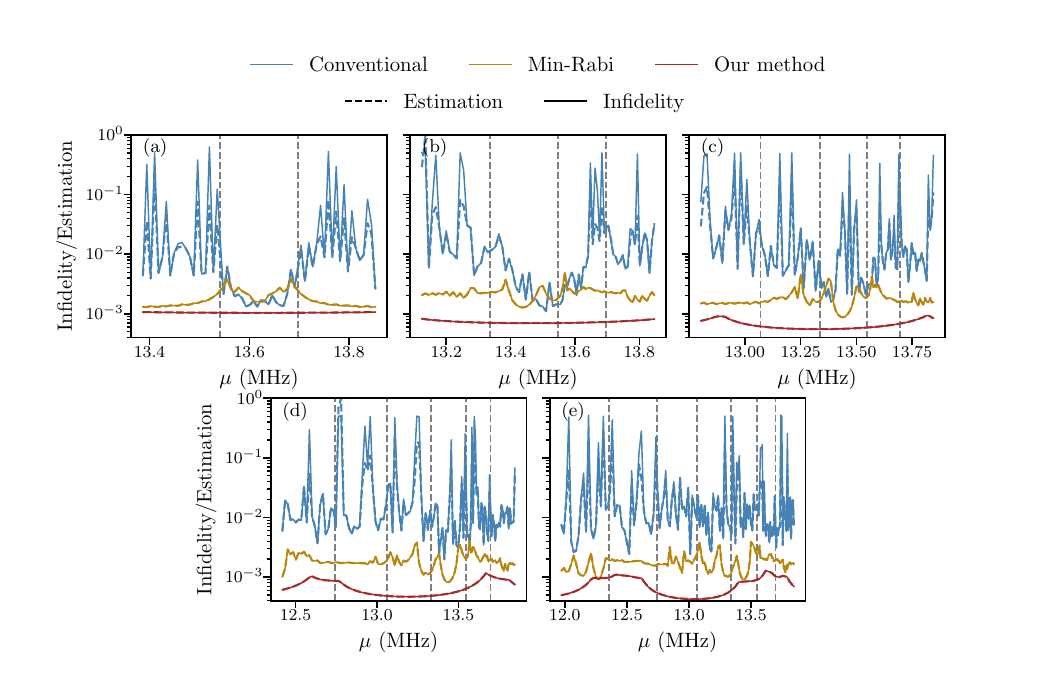}
    \caption{Infidelity versus laser detuning $\mu$. Panels (a)–(e) correspond to ion numbers $N=2,3,4,5,6$, respectively. The dashed gray vertical lines mark the phonon mode frequencies, which partition the detuning range into $N+1$ intervals. Each interval is uniformly divided into 19 subintervals, and simulations are performed at the 20 resulting endpoints. Simulations are carried out with $\Gamma_m^{\uparrow}=\Gamma_m^{\downarrow}=100\,\text{phonon/s}$ and gate duration $\tau=150\,\mu\text{s}$. In all subfigures, the red curves show the results obtained using the waveform optimized by our method [Eq.~\eqref{eq:optprob}], the blue curves correspond to the baseline waveform without error optimization, and the yellow curves represent the waveform optimized by minimizing the $L^2$-norm of the Rabi frequency. For each method, solid lines denote numerically simulated infidelities, while dashed lines indicate the corresponding theoretical estimates from Eq.~\eqref{eq:estimation}. The close overlap between dashed and solid lines in most cases demonstrates the accuracy of the theoretical estimation.}
    \label{fig:loginfid}
\end{figure*}

\begin{figure*}[t]
    \centering
    \includegraphics[width=\linewidth]{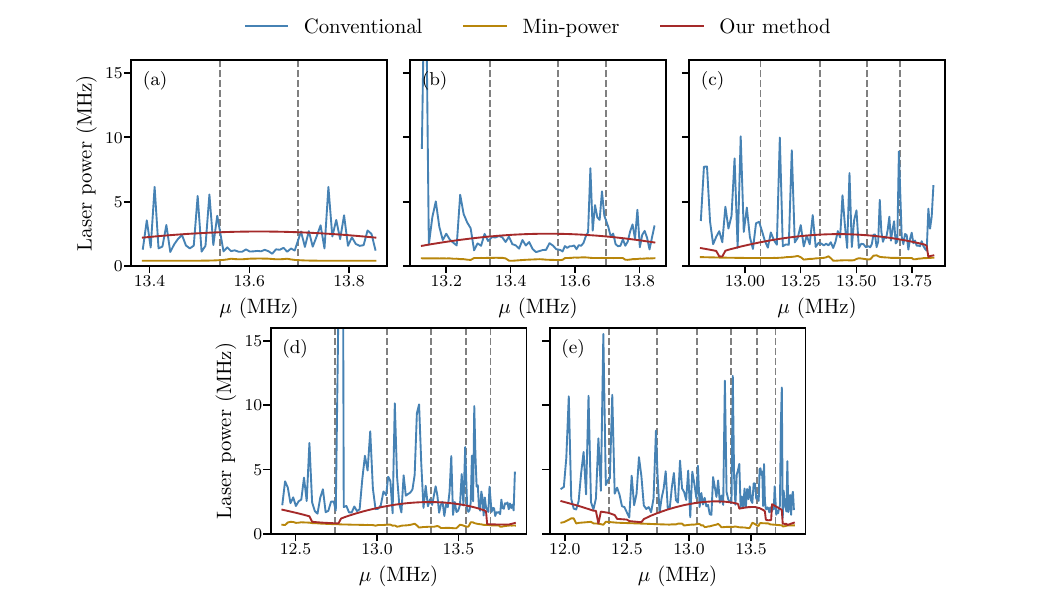}
    \caption{$L^2$-norm of Rabi frequencies versus laser detuning $\mu$. Panels (a)–(e) correspond to ion numbers $N=2,3,4,5,6$, respectively. The dashed gray vertical lines mark the phonon mode frequencies, which partition the detuning range into $N+1$ intervals. Each interval is uniformly divided into 19 subintervals, and simulations are performed at the 20 resulting endpoints. Simulations are carried out with $\Gamma_m^{\uparrow}=\Gamma_m^{\downarrow}=100\,\text{phonon/s}$ and gate duration $\tau=150\,\mu\text{s}$. In all subfigures, the red curves show the results obtained using the waveform optimized by our method [Eq.~\eqref{eq:optprob}], the blue curves correspond to the baseline waveform without error optimization, and the yellow curves represent the waveform optimized by minimizing the $L^2$-norm of the Rabi frequency.}
    \label{fig:l2norm}
\end{figure*}
In Fig.~\ref{fig:avg_infid}, we demonstrate that our method, which optimizes the heating error estimation from Eq.~\eqref{eq:estimation}, can effectively reduce heating error. However, the simulated infidelities do not exactly match the estimates. In this appendix, we provide a detailed comparison between the simulated infidelities and the error estimates. Additionally, we will show that the fluctuations in the waveforms generated by the conventional method lead to corresponding fluctuations in the infidelities.

Fig.~\ref{fig:loginfid} shows the infidelities and their corresponding estimations for three different methods, similar to Fig.~\ref{fig:avg_infid}(b), with ion numbers $N = 2, 3, 4, 5, 6$.
% In this figure, it is clear that, although the infidelities do not exactly match the estimates, they follow the same general trend, particularly for the min-Rabi method and our method. In fact, the estimations are consistently about four times larger than the infidelities for these two methods. This factor fluctuates around 0.01, remaining very stable. For the conventional method, this factor stays around 4 for most cases. However, as the fidelity improves, the gap between the estimates and the infidelities narrows. Although the exact origin of this near-constant factor remains unclear,
From this figure, we observe that the estimation agrees well with the numerically simulated infidelities, particularly for the min-Rabi method and our method. For the conventional method, although noticeable deviations appear, the overall trend is still well captured.
These deviations are likely attributable to the first-order approximation employed in the derivation of the estimation. They occur only at peaks of the infidelity, where higher-order contributions are no longer negligible.
These demonstrate that the optimization target Eq.~\eqref{eq:estimation} we’ve proposed provides a highly accurate description of the heating error. This consistency further supports the reliability of our error estimation method.

Fig.~\ref{fig:l2norm} shows the $L_2$-norm of Rabi frequencies for three different methods, similar to Fig.~\ref{fig:avg_infid}(c), with ion numbers $N = 2, 3, 4, 5, 6$. It is evident that the waveform produced by the min-Rabi method achieves a lower $L_2$ norm. Although our method does not minimize it to the same extent, it remains within a reasonable range and, importantly, is much more stable—avoiding the drastic fluctuations seen with the conventional method. Furthermore, by comparing Fig.~\ref{fig:loginfid} with Fig.~\ref{fig:l2norm}, it becomes clear that when the infidelity of the waveform from the conventional method exhibits an unreasonable peak, the $L_2$ norm of that waveform similarly shows an unreasonable peak. This pattern aligns well with the heating error estimate in Eq.~\eqref{eq:bound}, which suggests that heating error is related to the second-order term of the Rabi frequency. At the same time, a peak in the $L_2$ norm typically indicates that the phase-space trajectory moves far from the origin, where higher-order contributions beyond the first-order approximation become non-negligible. This explains why the infidelities exceed the estimations in Fig.~\ref{fig:loginfid}.

\section{Detuning error suppression}
\label{app:detuning}
\begin{figure*}[t]
    \center
    \includegraphics[width=0.95\linewidth]{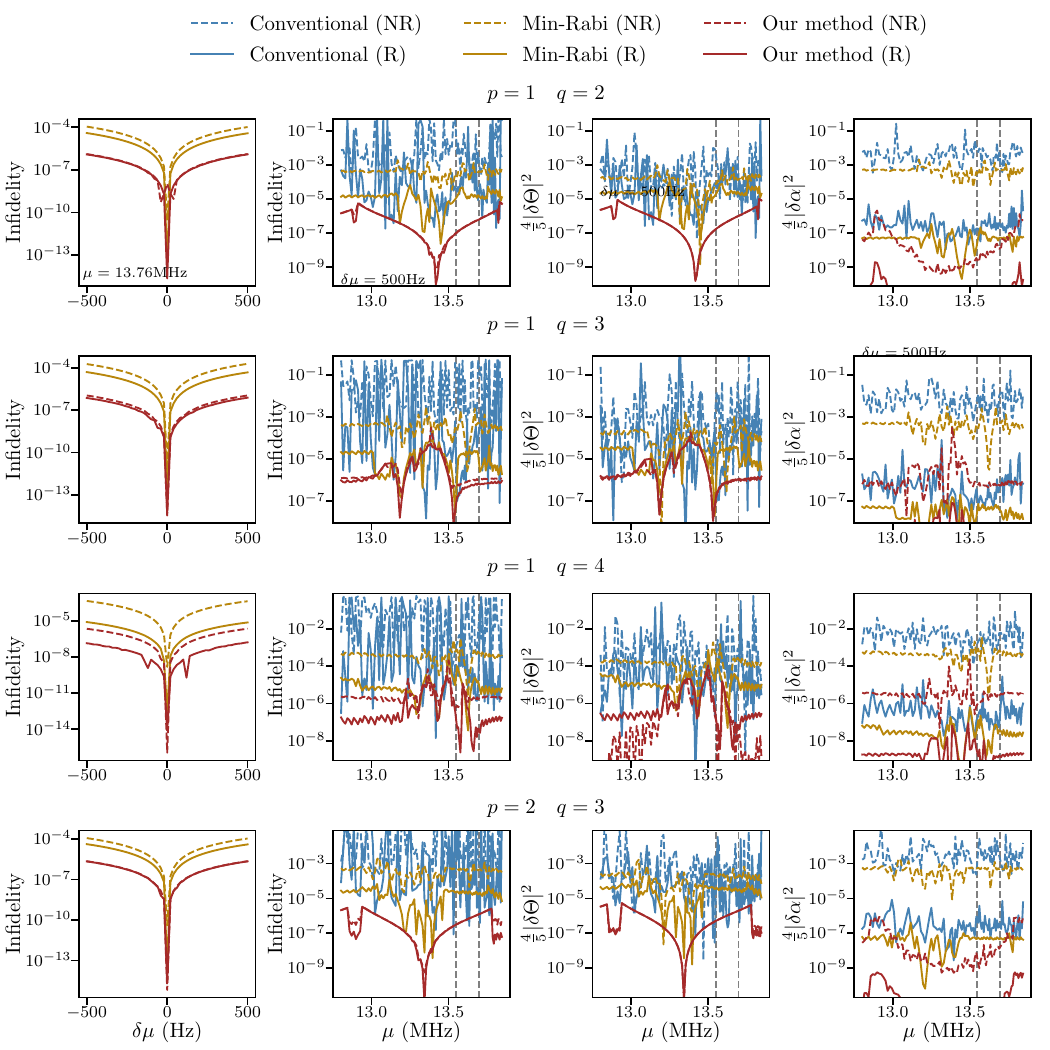}
    \caption{Infidelity and fluctuations versus detuning error $\delta \mu$ for all pairs $p,q$ of ions among $N=4$ ions. First column: Infidelity under various detuning errors $\delta \mu$ at a nominal detuning of $\mu = 13.76\,\text{MHz}$. Second to fourth columns: Infidelities, rotation-angle fluctuations $\delta\Theta$, and residual displacements $|\delta\alpha|^{2}$ versus various detuning $\mu$ under a fixed detuning error of $\delta \mu=500\,\text{Hz}$. In all subfigures, blue curves correspond to conventional method, yellow curves to the min-Rabi baseline and red curves to our method; NR denotes the non-robust versions (dashed), and R denotes the robust versions (solid) incorporating $\alpha$-robustness constraints. The two dashed vertical lines locate the two highest mode frequencies, i.e. the COM mode and the breathing mode. Here, $\delta\Theta$ denotes the deviation of the accumulated rotation angle from its target value, and $|\delta\alpha|^{2}$ denotes the total residual motional displacement $\sum_{j,m}|\alpha_j^{m}(\tau)|^{2}$. In (a)-(c), the NR and R versions of our method (dashed and solid red lines) nearly overlap, indicating that our method inherently provides robustness against detuning errors even without explicit constraints.}
    \label{fig:detuning_all}
\end{figure*}
In this appendix, we analyze the impact of detuning errors $\delta\mu$ on the gate fidelity of the Mølmer–Sørensen (MS) gate in Eq.~\eqref{eq:Utaum}. We theoretically derive the sensitivity of both residual phonon displacements and  the accumulated rotation angle, which are the two main contributors to gate infidelity in the presence of detuning errors. We then provide supplemental numerical results in Fig.~\ref{fig:detuning_all}.

Suppose that the initial phonon state is a separable thermal state with temperature $T$ and only two ions $p$ and $q$ are illuminated, according to Ref.~\cite{wu2018noise}, the average gate fidelity of Eq.~\eqref{eq:Utau} compared to the ideal gate $U_{\text{ideal}}=\exp\left(i\frac{\pi}{4}\sigma_p\sigma_q\right)$, at the leading order,  is given by
\begin{equation}
    \begin{aligned}
        \bar{F}\approx\frac{1}{10}&\left[4+2\Gamma_p\sin\left(2\Theta+\epsilon\right)\right. \\
        &\left.+2\Gamma_q\sin\left(2\Theta-\epsilon\right)+\Gamma_++\Gamma_-\right]
    \end{aligned}
\end{equation}
with
\begin{gather}
    \Gamma_j=\exp\left[-2\sum_{m}\left|\alpha_j^m(\tau)\right|^2\coth\left(\frac{\hbar\omega_m}{2k_BT}\right)\right], \\
    \Gamma_{\pm}=\exp\left[-2\sum_{m}\left|\alpha_p^m(\tau)\pm\alpha_q^m(\tau)\right|^2\coth\left(\frac{\hbar\omega_m}{2k_BT}\right)\right], \\
    \epsilon=2\sum_m\text{Im}\left(\alpha_p^m(\tau)\alpha_q^{m*}(\tau)\right).
\end{gather}
When there is a small detuning error $\delta\mu$, i.e., the actual detuning is $\mu+\delta\mu$, we can expand $\alpha_j^m(\tau)$ and $\Theta$ to the first order of $\delta\mu$ as
\begin{gather}
    \alpha_j^m(\tau)=0+\delta\alpha_j^m\approx\frac{\partial\alpha_j^m(\tau)}{\partial\mu}\delta\mu, \\
    \Theta=\frac{\pi}{4}+\delta\Theta\approx\frac{\pi}{4}+\frac{\partial\Theta}{\partial\mu}\delta\mu.
\end{gather}
Substituting these expansions into the expression of $\bar{F}$ and keeping only the first-order terms, we obtain
% \begin{equation}\label{eq:infdmu_}
%     \begin{aligned}
%         1-\bar{F}\approx& 8\left[\sum_{m}\left|\delta\alpha_p^m\right|^2 + \sum_{m}\left|\delta\alpha_q^m\right|^2\right]\coth\left(\frac{\hbar\omega_m}{2k_BT}\right)+2|\delta\Theta|^2\\ %+2\sum_m\text{Im}\left(\delta\alpha_p^m\delta\alpha_q^{m*}\right).
%         =&8|\delta\alpha|^2+2|\delta\Theta|^2.
%     \end{aligned}
% \end{equation}
\begin{equation}\label{eq:infdmu_}
    \begin{aligned}
        1-\bar{F}\approx& \frac{4}{5}\sum_{m}\left[\left|\delta\alpha_p^m\right|^2 + \left|\delta\alpha_q^m\right|^2\right]\coth\left(\frac{\hbar\omega_m}{2k_BT}\right)+\frac{4}{5}|\delta\Theta|^2\\ =&\frac{4}{5}\left[|\delta\alpha|^2+|\delta\Theta|^2\right].
    \end{aligned}
\end{equation}
This expression quantifies the gate infidelity arising from small detuning errors, highlighting the equal importance of minimizing both the sensitivity of phonon displacements and the rotation angle to detuning variations.

% \subsection{residual motional displacement $|\delta\alpha|^2$}
First, we compute the derivative of $\alpha_j^m(t)$ with respect to $\mu$:
\begin{align}
    \frac{\partial}{\partial\mu}\alpha_j^m(t)&=\frac{\eta_m b_j^m}{2}\int_0^t\Omega_j(t')\frac{\partial}{\partial\mu}e^{i(\omega_m-\mu)t'}dt' \\
    &=-\frac{i\eta_m b_j^m}{2}\int_0^t\Omega_j(t')t'e^{i(\omega_m-\mu)t'}dt' \\
    &=-i t \alpha_j^m(t)+i\int_0^t\alpha_j^m(t')dt' \\
    &=-i t \alpha_j^m(t)+it \bar{\alpha}_j^m(t),
\end{align}
where we have defined the time-averaged phonon displacement as $\bar{\alpha}_j^m(t)=\frac{1}{t}\int_0^t\alpha_j^m(t')dt'$. Thus, for $\alpha_j^m(\tau)=0$, we have
\begin{equation}
    \delta\alpha_j^m = \frac{\partial\alpha_j^m}{\partial\mu}\delta\mu = i\tau \bar{\alpha}_j^m(\tau)\delta\mu
\end{equation}

% \subsection{accumulated rotation angle error $\delta\Theta$}
Next, we compute the derivative of $\Theta$ with respect to $\mu$. We rewrite $\Theta$ as a function of $\alpha_j^m(t)$ according to~\cite{martinez2022analytical,ruzic2022frequency}:
\begin{equation}
    \Theta=\sum_m\int_0^\tau dt \text{Im}\left(\frac{\partial\alpha_p^m(t)}{\partial t}\alpha_q^{m*}(t)+\frac{\partial\alpha_q^m(t)}{\partial t}\alpha_p^{m*}(t)\right)
    \label{eq:theta_alpha}
\end{equation}
Taking the derivative with respect to $\mu$ for the first term inside the integral, we have
\begin{align}
    &\frac{\partial}{\partial\mu}\left(\frac{\partial\alpha_p^m(t)}{\partial t}\alpha_q^{m*}(t)\right) \\=&\frac{\partial\alpha_p^m(t)}{\partial t}\frac{\partial\alpha_q^{m*}(t)}{\partial\mu}+\frac{\partial}{\partial\mu}\left(\frac{\partial\alpha_p^m(t)}{\partial t}\right)\alpha_q^{m*}(t) \\
    =&it\frac{\partial\alpha_p^m(t)}{\partial t}\left(\alpha_q^{m*}(t)-\bar{\alpha}_q^{m*}(t)\right)-it\frac{\partial\alpha_p^m(t)}{\partial t}\alpha_q^{m*}(t) \\
    =&-it\frac{\partial\alpha_p^m(t)}{\partial t}\bar{\alpha}_q^{m*}(t)
\end{align}
Using integration by parts, we obtain
\begin{align}
    &\int_0^\tau dt \frac{\partial}{\partial\mu}\left(\frac{\partial\alpha_p^m(t)}{\partial t}\alpha_q^{m*}(t)\right) \\
    =&-i\int_0^\tau dt \, t \frac{\partial\alpha_p^m(t)}{\partial t}\bar{\alpha}_q^{m*}(t) \\
    =&-i\left.t\alpha_p^m(t)\bar{\alpha}_q^{m*}(t)\right|_0^\tau + i\int_0^\tau dt \, \alpha_p^m(t)\alpha_q^{m*}(t) \\
    =&i\int_0^\tau dt \, \alpha_p^m(t)\alpha_q^{m*}(t)
\end{align}
where we have used the boundary condition $\alpha_j^m(\tau)=0$ and the definition of $\bar{\alpha}_j^m(t)$. For all terms in Eq.~\eqref{eq:theta_alpha}, the calculation is similar, yielding the final result:
\begin{equation}
    \frac{\partial\Theta}{\partial\mu} = \text{Re}\left[\sum_m \int_0^\tau dt \, \alpha_p^m(t)\alpha_q^{m*}(t)+\int_0^\tau dt \, \alpha_q^m(t)\alpha_p^{m*}(t)\right]
\end{equation}
This result is consistent with the bound of the heating error in Eq.~\eqref{eq:bound}, i.e.
\begin{equation}
    \frac{\partial\Theta}{\partial\mu} \sim \Gamma^{-1}\text{Re}\left[E_{p,q}+E_{q,p}\right],
\end{equation}
where $\max_{m=1}^N\Gamma_m^{\uparrow}+\Gamma_m^{\downarrow}$. Since our method minimizes the heating error bound, it also effectively reduces the sensitivity of the rotation angle to detuning errors.

In Fig.~\ref{fig:detuning_all}, we compare the sensitivity to detuning errors of our method with that of the baseline approach for all ion pairs in a system of $N=4$ ions. In addition to the non-robust implementations, we also combine all three methods with the robustness constraints introduced in Ref.~\cite{leung2018robust}, which enforce $\partial \alpha / \partial \mu = 0$; these are denoted as the robust versions. For each ion pair and over most values of the detuning $\mu$, our method—both with and without the $\alpha$-robustness constraints—exhibits smaller fluctuations $|\delta\Theta|^2$ in the accumulated rotation angle, and consequently lower infidelities under detuning errors. This demonstrates that minimizing the heating-error bound inherently enhances robustness against detuning fluctuations over a broad parameter range, even without explicitly imposing the $\alpha$-robustness constraints of Ref.~\cite{leung2018robust}.

Moreover, even the non-robust version of our method tends to yield relatively small residual motional displacements $|\delta\alpha|^{2}$ for most parameter regimes. This behavior can be attributed to the fact that the optimized waveforms typically drive the displacement $\alpha(t)$ to oscillate around the origin (see Fig.~\ref{fig:alpha}), resulting in a reduced average displacement $\bar{\alpha}$. %\red{This mechanism is similar to that discussed in Ref.\cite{hayes2012coherent}, where Walsh-1 waveforms are employed to suppress $|\delta\alpha|^2$.}

Below, we briefly outline the derivation. We assume that the Rabi frequency varies slowly in time while its sign alternates periodically, i.e.,
\begin{equation}
    \Omega(t)=\begin{cases}
        (-1)^{0}|c_1| \quad &0\leq t< \tau/L \\
        (-1)^{1}|c_2| \quad &\tau/L\leq t<2\tau/L \\
        \quad\vdots \\
        (-1)^{L-1}|c_L| \quad &(L-1)\tau/L\leq t\leq \tau
    \end{cases},
\end{equation}
where $|c_l| \approx |c_{l+1}|$ describes a slowly varying envelope. In the presence of a detuning error $\delta\mu$, the motional displacement, according to Eq.~\eqref{eq:alpha}, can be written as
\begin{align}
    \alpha(\tau)\propto\int_0^\tau \Omega(t)e^{i\tilde{\Delta}t}\,dt,
\end{align}
where $\tilde{\Delta}=\Delta-\delta\mu$, and the mode and ion indices have been omitted for simplicity. Substituting $\Omega(t)$, we can obtain
\begin{align}
    \alpha(\tau)\propto&\sum_{l=1}^L(-1)^{l-1}\int_{\frac{l-1}{L}\tau}^{\frac{l}{L}\tau}|c_l|e^{i\tilde{\Delta}t}dt \\
    =&\sum_{l=1}^{L/2}\int_{\frac{2l-2}{L}\tau}^{\frac{2l-1}{L}\tau}|c_{2l-1}|e^{i\tilde{\Delta}t}dt-\sum_{l=1}^{L/2}\int_{\frac{2l-1}{L}\tau}^{\frac{2l}{L}\tau}|c_{2l}|e^{i\tilde{\Delta}t}dt \\
    \approx&\sum_{l=1}^{L/2}\left(1-e^{i\tilde{\Delta}\tau/L}\right)\int_{\frac{2l-2}{L}\tau}^{\frac{2l-1}{L}\tau}|c_{2l-1}|e^{i\tilde{\Delta}t}dt \\
    \approx&\frac{-i\tilde{\Delta}\tau}{2L}\int_0^\tau|\Omega(t)|e^{i\tilde{\Delta}t}dt,
\end{align}
where in the second step we have used $|c_{2l}|\approx |c_{2l-1}|$.
Thus, the residual displacement is suppressed to $O(\tilde{\Delta}\tau/L)$, even in the presence of detuning errors.
\begin{figure}[t]
    \center
    \includegraphics[width=\columnwidth]{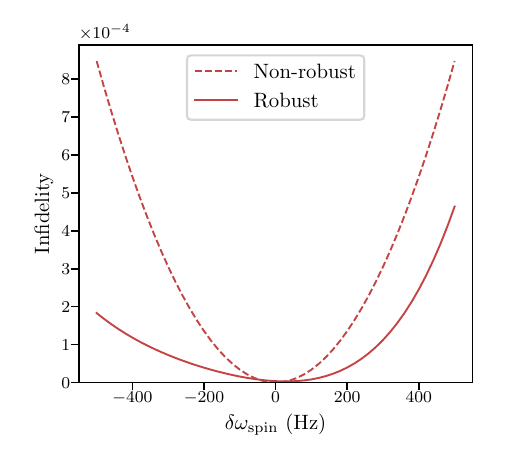}
    \caption{Infidelity versus spin frequency error $\delta \omega_{\text{spin}}$ for a system of $N=3$ ions, targeting the left-most two ions. The dashed line corresponds to our method without spin frequency-error robustness, while the solid line shows our method combined with the  spin frequency-error robustness technique introduced in Ref.~\cite{Zhang.25}.}
    \label{fig:asym}
\end{figure}

\section{Combination with other error suppression methods}
\label{app:combine}
\begin{figure}[t]
    \center
    \includegraphics[width=\columnwidth]{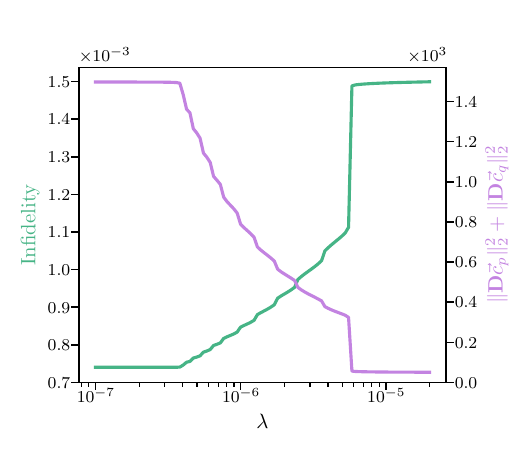}
    \caption{Infidelity and Rabi-frequency smoothness as functions of the optimization weighting factor $\lambda$ for a system of N=3 ions, targeting the left-most two ions. The green curve shows the infidelity, while the purple curve denotes the smoothness of the Rabi frequency. The smoothness is quantified by the squared finite-difference norm $\|\mathbf{D}\vec{c}\|_2^2 = \sum_{l=1}^{L-1}(c_{l+1}-c_l)^2$, where $\vec{c}=(c_1,c_2,\ldots,c_L)$ is the coefficient vector of the Rabi-frequency waveform and $\mathbf{D}$ is the finite-difference matrix.}
    \label{fig:smooth}
\end{figure}
In this appendix, we demonstrate that our method for minimizing heating errors can be effectively combined with other error suppression techniques, such as spin-frequency error robustness~\cite{Zhang.25} and amplitude smoothing~\cite{cheng2023robust}.

Spin-frequency errors introduce an additional spin-rotation term in the Hamiltonian,
\begin{equation}
H’(t)=\sum_{j=1}^N \delta\omega_{\text{spin}}\,\sigma_j^z + H(t),
\end{equation}
where $H(t)$ denotes the noiseless Hamiltonian defined in Eq.~\eqref{eq:H}, and $\delta\omega_{\text{spin}}$ characterizes the magnitude of the spin-frequency error. We combine our heating-error minimization framework with the spin-frequency-error robustness technique proposed in Ref.~\cite{Zhang.25}, which employs three MS gates to dynamically decouple this error source. Fig.~\ref{fig:asym} shows the resulting infidelity as a function of $\delta\omega_{\text{spin}}$ for a system of N=3 ions, targeting the left-most two ions. The results demonstrate that this combined approach substantially suppresses infidelity over a broad range of spin-frequency errors.

We also combine our method with amplitude smoothing techniques~\cite{cheng2023robust} to generate smooth Rabi frequency profiles. We add a penalty term to our optimization objective that minimizes the squared derivative of the Rabi frequency, i.e.
\begin{equation}
    \minimize \quad E + \lambda\left[\|\mathbf{D}\vec{c}_p\|_2^2 + \|\mathbf{D}\vec{c}_q\|_2^2\right],
\end{equation}
where $\mathbf{D}$ is the finite difference matrix defined as
\begin{equation}
    \mathbf{D}=\begin{bmatrix}
        1 & -1 & 0 & \cdots & 0 \\
        0 & 1 & -1 & \cdots & 0 \\
        \vdots & \vdots & \vdots & \ddots & \vdots \\
        0 & 0 & 0 & \cdots & -1 \\
        0 & 0 & 0 & \cdots & 1
    \end{bmatrix},
\end{equation}
and $\lambda$ is a weighting factor that controls the trade-off between heating error minimization and waveform smoothness. In Fig.~\ref{fig:smooth}, the results demonstrate that this approach effectively reduces high-frequency components in the Rabi frequency profile while still achieving low infidelity. By adjusting $\lambda$, we only need to increase the heating error from $7.4\times10^{-4}$ to $1.5\times10^{-3}$ to reduce the difference of the waveform by 30 times. Therefore, we can generate Rabi frequency profiles that are both low in heating error and smooth, making them more feasible for experimental implementation.

\end{appendix}

\bibliographystyle{apsrev4-2}
\bibliography{ref}

\begin{thebibliography}{37}%
\makeatletter
\providecommand \@ifxundefined [1]{%
 \@ifx{#1\undefined}
}%
\providecommand \@ifnum [1]{%
 \ifnum #1\expandafter \@firstoftwo
 \else \expandafter \@secondoftwo
 \fi
}%
\providecommand \@ifx [1]{%
 \ifx #1\expandafter \@firstoftwo
 \else \expandafter \@secondoftwo
 \fi
}%
\providecommand \natexlab [1]{#1}%
\providecommand \enquote  [1]{``#1''}%
\providecommand \bibnamefont  [1]{#1}%
\providecommand \bibfnamefont [1]{#1}%
\providecommand \citenamefont [1]{#1}%
\providecommand \href@noop [0]{\@secondoftwo}%
\providecommand \href [0]{\begingroup \@sanitize@url \@href}%
\providecommand \@href[1]{\@@startlink{#1}\@@href}%
\providecommand \@@href[1]{\endgroup#1\@@endlink}%
\providecommand \@sanitize@url [0]{\catcode `\\12\catcode `\$12\catcode
  `\&12\catcode `\#12\catcode `\^12\catcode `\_12\catcode `\%12\relax}%
\providecommand \@@startlink[1]{}%
\providecommand \@@endlink[0]{}%
\providecommand \url  [0]{\begingroup\@sanitize@url \@url }%
\providecommand \@url [1]{\endgroup\@href {#1}{\urlprefix }}%
\providecommand \urlprefix  [0]{URL }%
\providecommand \Eprint [0]{\href }%
\providecommand \doibase [0]{https://doi.org/}%
\providecommand \selectlanguage [0]{\@gobble}%
\providecommand \bibinfo  [0]{\@secondoftwo}%
\providecommand \bibfield  [0]{\@secondoftwo}%
\providecommand \translation [1]{[#1]}%
\providecommand \BibitemOpen [0]{}%
\providecommand \bibitemStop [0]{}%
\providecommand \bibitemNoStop [0]{.\EOS\space}%
\providecommand \EOS [0]{\spacefactor3000\relax}%
\providecommand \BibitemShut  [1]{\csname bibitem#1\endcsname}%
\let\auto@bib@innerbib\@empty
%</preamble>
\bibitem [{\citenamefont {H{\"a}ffner}\ \emph {et~al.}(2008)\citenamefont
  {H{\"a}ffner}, \citenamefont {Roos},\ and\ \citenamefont
  {Blatt}}]{haffner2008quantum}%
  \BibitemOpen
  \bibfield  {author} {\bibinfo {author} {\bibfnamefont {H.}~\bibnamefont
  {H{\"a}ffner}}, \bibinfo {author} {\bibfnamefont {C.~F.}\ \bibnamefont
  {Roos}},\ and\ \bibinfo {author} {\bibfnamefont {R.}~\bibnamefont {Blatt}},\
  }\href@noop {} {\bibfield  {journal} {\bibinfo  {journal} {Physics reports}\
  }\textbf {\bibinfo {volume} {469}},\ \bibinfo {pages} {155} (\bibinfo {year}
  {2008})}\BibitemShut {NoStop}%
\bibitem [{\citenamefont {Monroe}\ and\ \citenamefont
  {Kim}(2013)}]{monroe2013scaling}%
  \BibitemOpen
  \bibfield  {author} {\bibinfo {author} {\bibfnamefont {C.}~\bibnamefont
  {Monroe}}\ and\ \bibinfo {author} {\bibfnamefont {J.}~\bibnamefont {Kim}},\
  }\href@noop {} {\bibfield  {journal} {\bibinfo  {journal} {Science}\ }\textbf
  {\bibinfo {volume} {339}},\ \bibinfo {pages} {1164} (\bibinfo {year}
  {2013})}\BibitemShut {NoStop}%
\bibitem [{\citenamefont {Bruzewicz}\ \emph {et~al.}(2019)\citenamefont
  {Bruzewicz}, \citenamefont {Chiaverini}, \citenamefont {McConnell},\ and\
  \citenamefont {Sage}}]{bruzewicz2019trapped}%
  \BibitemOpen
  \bibfield  {author} {\bibinfo {author} {\bibfnamefont {C.~D.}\ \bibnamefont
  {Bruzewicz}}, \bibinfo {author} {\bibfnamefont {J.}~\bibnamefont
  {Chiaverini}}, \bibinfo {author} {\bibfnamefont {R.}~\bibnamefont
  {McConnell}},\ and\ \bibinfo {author} {\bibfnamefont {J.~M.}\ \bibnamefont
  {Sage}},\ }\href@noop {} {\bibfield  {journal} {\bibinfo  {journal} {Applied
  physics reviews}\ }\textbf {\bibinfo {volume} {6}} (\bibinfo {year}
  {2019})}\BibitemShut {NoStop}%
\bibitem [{\citenamefont {Harty}\ \emph {et~al.}(2014)\citenamefont {Harty},
  \citenamefont {Allcock}, \citenamefont {Ballance}, \citenamefont {Guidoni},
  \citenamefont {Janacek}, \citenamefont {Linke}, \citenamefont {Stacey},\ and\
  \citenamefont {Lucas}}]{harty2014high}%
  \BibitemOpen
  \bibfield  {author} {\bibinfo {author} {\bibfnamefont {T.}~\bibnamefont
  {Harty}}, \bibinfo {author} {\bibfnamefont {D.}~\bibnamefont {Allcock}},
  \bibinfo {author} {\bibfnamefont {C.~J.~.}\ \bibnamefont {Ballance}},
  \bibinfo {author} {\bibfnamefont {L.}~\bibnamefont {Guidoni}}, \bibinfo
  {author} {\bibfnamefont {H.}~\bibnamefont {Janacek}}, \bibinfo {author}
  {\bibfnamefont {N.}~\bibnamefont {Linke}}, \bibinfo {author} {\bibfnamefont
  {D.}~\bibnamefont {Stacey}},\ and\ \bibinfo {author} {\bibfnamefont
  {D.}~\bibnamefont {Lucas}},\ }\href@noop {} {\bibfield  {journal} {\bibinfo
  {journal} {Physical review letters}\ }\textbf {\bibinfo {volume} {113}},\
  \bibinfo {pages} {220501} (\bibinfo {year} {2014})}\BibitemShut {NoStop}%
\bibitem [{\citenamefont {Moses}\ \emph {et~al.}(2023)\citenamefont {Moses},
  \citenamefont {Baldwin}, \citenamefont {Allman}, \citenamefont {Ancona},
  \citenamefont {Ascarrunz}, \citenamefont {Barnes}, \citenamefont
  {Bartolotta}, \citenamefont {Bjork}, \citenamefont {Blanchard}, \citenamefont
  {Bohn} \emph {et~al.}}]{moses2023race}%
  \BibitemOpen
  \bibfield  {author} {\bibinfo {author} {\bibfnamefont {S.~A.}\ \bibnamefont
  {Moses}}, \bibinfo {author} {\bibfnamefont {C.~H.}\ \bibnamefont {Baldwin}},
  \bibinfo {author} {\bibfnamefont {M.~S.}\ \bibnamefont {Allman}}, \bibinfo
  {author} {\bibfnamefont {R.}~\bibnamefont {Ancona}}, \bibinfo {author}
  {\bibfnamefont {L.}~\bibnamefont {Ascarrunz}}, \bibinfo {author}
  {\bibfnamefont {C.}~\bibnamefont {Barnes}}, \bibinfo {author} {\bibfnamefont
  {J.}~\bibnamefont {Bartolotta}}, \bibinfo {author} {\bibfnamefont
  {B.}~\bibnamefont {Bjork}}, \bibinfo {author} {\bibfnamefont
  {P.}~\bibnamefont {Blanchard}}, \bibinfo {author} {\bibfnamefont
  {M.}~\bibnamefont {Bohn}}, \emph {et~al.},\ }\href@noop {} {\bibfield
  {journal} {\bibinfo  {journal} {Physical Review X}\ }\textbf {\bibinfo
  {volume} {13}},\ \bibinfo {pages} {041052} (\bibinfo {year}
  {2023})}\BibitemShut {NoStop}%
\bibitem [{\citenamefont {DeCross}\ \emph {et~al.}(2024)\citenamefont
  {DeCross}, \citenamefont {Haghshenas}, \citenamefont {Liu}, \citenamefont
  {Alexeev}, \citenamefont {Baldwin}, \citenamefont {Bartolotta}, \citenamefont
  {Bohn}, \citenamefont {Chertkov}, \citenamefont {Colina}, \citenamefont
  {DelVento} \emph {et~al.}}]{decross2024computational}%
  \BibitemOpen
  \bibfield  {author} {\bibinfo {author} {\bibfnamefont {M.}~\bibnamefont
  {DeCross}}, \bibinfo {author} {\bibfnamefont {R.}~\bibnamefont {Haghshenas}},
  \bibinfo {author} {\bibfnamefont {M.}~\bibnamefont {Liu}}, \bibinfo {author}
  {\bibfnamefont {Y.}~\bibnamefont {Alexeev}}, \bibinfo {author} {\bibfnamefont
  {C.~H.}\ \bibnamefont {Baldwin}}, \bibinfo {author} {\bibfnamefont {J.~P.}\
  \bibnamefont {Bartolotta}}, \bibinfo {author} {\bibfnamefont
  {M.}~\bibnamefont {Bohn}}, \bibinfo {author} {\bibfnamefont {E.}~\bibnamefont
  {Chertkov}}, \bibinfo {author} {\bibfnamefont {J.}~\bibnamefont {Colina}},
  \bibinfo {author} {\bibfnamefont {D.}~\bibnamefont {DelVento}}, \emph
  {et~al.},\ }\href@noop {} {\bibfield  {journal} {\bibinfo  {journal} {arXiv
  preprint arXiv:2406.02501}\ } (\bibinfo {year} {2024})}\BibitemShut {NoStop}%
\bibitem [{\citenamefont {Guo}\ \emph {et~al.}(2024)\citenamefont {Guo},
  \citenamefont {Wu}, \citenamefont {Ye}, \citenamefont {Zhang}, \citenamefont
  {Lian}, \citenamefont {Yao}, \citenamefont {Wang}, \citenamefont {Yan},
  \citenamefont {Yi}, \citenamefont {Xu} \emph {et~al.}}]{guo2024site}%
  \BibitemOpen
  \bibfield  {author} {\bibinfo {author} {\bibfnamefont {S.-A.}\ \bibnamefont
  {Guo}}, \bibinfo {author} {\bibfnamefont {Y.-K.}\ \bibnamefont {Wu}},
  \bibinfo {author} {\bibfnamefont {J.}~\bibnamefont {Ye}}, \bibinfo {author}
  {\bibfnamefont {L.}~\bibnamefont {Zhang}}, \bibinfo {author} {\bibfnamefont
  {W.-Q.}\ \bibnamefont {Lian}}, \bibinfo {author} {\bibfnamefont
  {R.}~\bibnamefont {Yao}}, \bibinfo {author} {\bibfnamefont {Y.}~\bibnamefont
  {Wang}}, \bibinfo {author} {\bibfnamefont {R.-Y.}\ \bibnamefont {Yan}},
  \bibinfo {author} {\bibfnamefont {Y.-J.}\ \bibnamefont {Yi}}, \bibinfo
  {author} {\bibfnamefont {Y.-L.}\ \bibnamefont {Xu}}, \emph {et~al.},\
  }\href@noop {} {\bibfield  {journal} {\bibinfo  {journal} {Nature}\ }\textbf
  {\bibinfo {volume} {630}},\ \bibinfo {pages} {613} (\bibinfo {year}
  {2024})}\BibitemShut {NoStop}%
\bibitem [{\citenamefont {Feng}\ \emph {et~al.}(2023)\citenamefont {Feng},
  \citenamefont {Katz}, \citenamefont {Haack}, \citenamefont {Maghrebi},
  \citenamefont {Gorshkov}, \citenamefont {Gong}, \citenamefont {Cetina},\ and\
  \citenamefont {Monroe}}]{feng2023continuous}%
  \BibitemOpen
  \bibfield  {author} {\bibinfo {author} {\bibfnamefont {L.}~\bibnamefont
  {Feng}}, \bibinfo {author} {\bibfnamefont {O.}~\bibnamefont {Katz}}, \bibinfo
  {author} {\bibfnamefont {C.}~\bibnamefont {Haack}}, \bibinfo {author}
  {\bibfnamefont {M.}~\bibnamefont {Maghrebi}}, \bibinfo {author}
  {\bibfnamefont {A.~V.}\ \bibnamefont {Gorshkov}}, \bibinfo {author}
  {\bibfnamefont {Z.}~\bibnamefont {Gong}}, \bibinfo {author} {\bibfnamefont
  {M.}~\bibnamefont {Cetina}},\ and\ \bibinfo {author} {\bibfnamefont
  {C.}~\bibnamefont {Monroe}},\ }\href@noop {} {\bibfield  {journal} {\bibinfo
  {journal} {Nature}\ }\textbf {\bibinfo {volume} {623}},\ \bibinfo {pages}
  {713} (\bibinfo {year} {2023})}\BibitemShut {NoStop}%
\bibitem [{\citenamefont {Iqbal}\ \emph {et~al.}(2024)\citenamefont {Iqbal},
  \citenamefont {Tantivasadakarn}, \citenamefont {Verresen}, \citenamefont
  {Campbell}, \citenamefont {Dreiling}, \citenamefont {Figgatt}, \citenamefont
  {Gaebler}, \citenamefont {Johansen}, \citenamefont {Mills}, \citenamefont
  {Moses} \emph {et~al.}}]{iqbal2024non}%
  \BibitemOpen
  \bibfield  {author} {\bibinfo {author} {\bibfnamefont {M.}~\bibnamefont
  {Iqbal}}, \bibinfo {author} {\bibfnamefont {N.}~\bibnamefont
  {Tantivasadakarn}}, \bibinfo {author} {\bibfnamefont {R.}~\bibnamefont
  {Verresen}}, \bibinfo {author} {\bibfnamefont {S.~L.}\ \bibnamefont
  {Campbell}}, \bibinfo {author} {\bibfnamefont {J.~M.}\ \bibnamefont
  {Dreiling}}, \bibinfo {author} {\bibfnamefont {C.}~\bibnamefont {Figgatt}},
  \bibinfo {author} {\bibfnamefont {J.~P.}\ \bibnamefont {Gaebler}}, \bibinfo
  {author} {\bibfnamefont {J.}~\bibnamefont {Johansen}}, \bibinfo {author}
  {\bibfnamefont {M.}~\bibnamefont {Mills}}, \bibinfo {author} {\bibfnamefont
  {S.~A.}\ \bibnamefont {Moses}}, \emph {et~al.},\ }\href@noop {} {\bibfield
  {journal} {\bibinfo  {journal} {Nature}\ }\textbf {\bibinfo {volume} {626}},\
  \bibinfo {pages} {505} (\bibinfo {year} {2024})}\BibitemShut {NoStop}%
\bibitem [{\citenamefont {Chertkov}\ \emph {et~al.}(2023)\citenamefont
  {Chertkov}, \citenamefont {Cheng}, \citenamefont {Potter}, \citenamefont
  {Gopalakrishnan}, \citenamefont {Gatterman}, \citenamefont {Gerber},
  \citenamefont {Gilmore}, \citenamefont {Gresh}, \citenamefont {Hall},
  \citenamefont {Hankin} \emph {et~al.}}]{chertkov2023characterizing}%
  \BibitemOpen
  \bibfield  {author} {\bibinfo {author} {\bibfnamefont {E.}~\bibnamefont
  {Chertkov}}, \bibinfo {author} {\bibfnamefont {Z.}~\bibnamefont {Cheng}},
  \bibinfo {author} {\bibfnamefont {A.~C.}\ \bibnamefont {Potter}}, \bibinfo
  {author} {\bibfnamefont {S.}~\bibnamefont {Gopalakrishnan}}, \bibinfo
  {author} {\bibfnamefont {T.~M.}\ \bibnamefont {Gatterman}}, \bibinfo {author}
  {\bibfnamefont {J.~A.}\ \bibnamefont {Gerber}}, \bibinfo {author}
  {\bibfnamefont {K.}~\bibnamefont {Gilmore}}, \bibinfo {author} {\bibfnamefont
  {D.}~\bibnamefont {Gresh}}, \bibinfo {author} {\bibfnamefont
  {A.}~\bibnamefont {Hall}}, \bibinfo {author} {\bibfnamefont {A.}~\bibnamefont
  {Hankin}}, \emph {et~al.},\ }\href@noop {} {\bibfield  {journal} {\bibinfo
  {journal} {Nature Physics}\ }\textbf {\bibinfo {volume} {19}},\ \bibinfo
  {pages} {1799} (\bibinfo {year} {2023})}\BibitemShut {NoStop}%
\bibitem [{\citenamefont {Ballance}\ \emph {et~al.}(2016)\citenamefont
  {Ballance}, \citenamefont {Harty}, \citenamefont {Linke}, \citenamefont
  {Sepiol},\ and\ \citenamefont {Lucas}}]{ballance2016high}%
  \BibitemOpen
  \bibfield  {author} {\bibinfo {author} {\bibfnamefont {C.~J.}\ \bibnamefont
  {Ballance}}, \bibinfo {author} {\bibfnamefont {T.~P.}\ \bibnamefont {Harty}},
  \bibinfo {author} {\bibfnamefont {N.~M.}\ \bibnamefont {Linke}}, \bibinfo
  {author} {\bibfnamefont {M.~A.}\ \bibnamefont {Sepiol}},\ and\ \bibinfo
  {author} {\bibfnamefont {D.~M.}\ \bibnamefont {Lucas}},\ }\href@noop {}
  {\bibfield  {journal} {\bibinfo  {journal} {Physical review letters}\
  }\textbf {\bibinfo {volume} {117}},\ \bibinfo {pages} {060504} (\bibinfo
  {year} {2016})}\BibitemShut {NoStop}%
\bibitem [{\citenamefont {Baldwin}\ \emph {et~al.}(2021)\citenamefont
  {Baldwin}, \citenamefont {Bjork}, \citenamefont {Foss-Feig}, \citenamefont
  {Gaebler}, \citenamefont {Hayes}, \citenamefont {Kokish}, \citenamefont
  {Langer}, \citenamefont {Sedlacek}, \citenamefont {Stack},\ and\
  \citenamefont {Vittorini}}]{baldwin2021high}%
  \BibitemOpen
  \bibfield  {author} {\bibinfo {author} {\bibfnamefont {C.}~\bibnamefont
  {Baldwin}}, \bibinfo {author} {\bibfnamefont {B.}~\bibnamefont {Bjork}},
  \bibinfo {author} {\bibfnamefont {M.}~\bibnamefont {Foss-Feig}}, \bibinfo
  {author} {\bibfnamefont {J.}~\bibnamefont {Gaebler}}, \bibinfo {author}
  {\bibfnamefont {D.}~\bibnamefont {Hayes}}, \bibinfo {author} {\bibfnamefont
  {M.}~\bibnamefont {Kokish}}, \bibinfo {author} {\bibfnamefont
  {C.}~\bibnamefont {Langer}}, \bibinfo {author} {\bibfnamefont
  {J.}~\bibnamefont {Sedlacek}}, \bibinfo {author} {\bibfnamefont
  {D.}~\bibnamefont {Stack}},\ and\ \bibinfo {author} {\bibfnamefont
  {G.}~\bibnamefont {Vittorini}},\ }\href@noop {} {\bibfield  {journal}
  {\bibinfo  {journal} {Physical Review A}\ }\textbf {\bibinfo {volume}
  {103}},\ \bibinfo {pages} {012603} (\bibinfo {year} {2021})}\BibitemShut
  {NoStop}%
\bibitem [{\citenamefont {Wu}\ \emph {et~al.}(2018)\citenamefont {Wu},
  \citenamefont {Wang},\ and\ \citenamefont {Duan}}]{wu2018noise}%
  \BibitemOpen
  \bibfield  {author} {\bibinfo {author} {\bibfnamefont {Y.}~\bibnamefont
  {Wu}}, \bibinfo {author} {\bibfnamefont {S.-T.}\ \bibnamefont {Wang}},\ and\
  \bibinfo {author} {\bibfnamefont {L.-M.}\ \bibnamefont {Duan}},\ }\href@noop
  {} {\bibfield  {journal} {\bibinfo  {journal} {Physical Review A}\ }\textbf
  {\bibinfo {volume} {97}},\ \bibinfo {pages} {062325} (\bibinfo {year}
  {2018})}\BibitemShut {NoStop}%
\bibitem [{\citenamefont {He}\ \emph {et~al.}(2024)\citenamefont {He},
  \citenamefont {Zhang}, \citenamefont {Yuan}, \citenamefont {Shen},\ and\
  \citenamefont {Zhang}}]{he2024scaling}%
  \BibitemOpen
  \bibfield  {author} {\bibinfo {author} {\bibfnamefont {W.}~\bibnamefont
  {He}}, \bibinfo {author} {\bibfnamefont {W.}~\bibnamefont {Zhang}}, \bibinfo
  {author} {\bibfnamefont {X.}~\bibnamefont {Yuan}}, \bibinfo {author}
  {\bibfnamefont {Y.}~\bibnamefont {Shen}},\ and\ \bibinfo {author}
  {\bibfnamefont {X.-M.}\ \bibnamefont {Zhang}},\ }\href@noop {} {\bibfield
  {journal} {\bibinfo  {journal} {Journal of Physics A: Mathematical and
  Theoretical}\ }\textbf {\bibinfo {volume} {57}},\ \bibinfo {pages} {375306}
  (\bibinfo {year} {2024})}\BibitemShut {NoStop}%
\bibitem [{\citenamefont {Haddadfarshi}\ and\ \citenamefont
  {Mintert}(2016)}]{Haddadfarshi_2016}%
  \BibitemOpen
  \bibfield  {author} {\bibinfo {author} {\bibfnamefont {F.}~\bibnamefont
  {Haddadfarshi}}\ and\ \bibinfo {author} {\bibfnamefont {F.}~\bibnamefont
  {Mintert}},\ }\href {https://doi.org/10.1088/1367-2630/18/12/123007}
  {\bibfield  {journal} {\bibinfo  {journal} {New Journal of Physics}\ }\textbf
  {\bibinfo {volume} {18}},\ \bibinfo {pages} {123007} (\bibinfo {year}
  {2016})}\BibitemShut {NoStop}%
\bibitem [{\citenamefont {Shapira}\ \emph {et~al.}(2018)\citenamefont
  {Shapira}, \citenamefont {Shaniv}, \citenamefont {Manovitz}, \citenamefont
  {Akerman},\ and\ \citenamefont {Ozeri}}]{shapira2018robust}%
  \BibitemOpen
  \bibfield  {author} {\bibinfo {author} {\bibfnamefont {Y.}~\bibnamefont
  {Shapira}}, \bibinfo {author} {\bibfnamefont {R.}~\bibnamefont {Shaniv}},
  \bibinfo {author} {\bibfnamefont {T.}~\bibnamefont {Manovitz}}, \bibinfo
  {author} {\bibfnamefont {N.}~\bibnamefont {Akerman}},\ and\ \bibinfo {author}
  {\bibfnamefont {R.}~\bibnamefont {Ozeri}},\ }\href@noop {} {\bibfield
  {journal} {\bibinfo  {journal} {Physical review letters}\ }\textbf {\bibinfo
  {volume} {121}},\ \bibinfo {pages} {180502} (\bibinfo {year}
  {2018})}\BibitemShut {NoStop}%
\bibitem [{\citenamefont {Webb}\ \emph {et~al.}(2018)\citenamefont {Webb},
  \citenamefont {Webster}, \citenamefont {Collingbourne}, \citenamefont
  {Bretaud}, \citenamefont {Lawrence}, \citenamefont {Weidt}, \citenamefont
  {Mintert},\ and\ \citenamefont {Hensinger}}]{webb2018resilient}%
  \BibitemOpen
  \bibfield  {author} {\bibinfo {author} {\bibfnamefont {A.~E.}\ \bibnamefont
  {Webb}}, \bibinfo {author} {\bibfnamefont {S.~C.}\ \bibnamefont {Webster}},
  \bibinfo {author} {\bibfnamefont {S.}~\bibnamefont {Collingbourne}}, \bibinfo
  {author} {\bibfnamefont {D.}~\bibnamefont {Bretaud}}, \bibinfo {author}
  {\bibfnamefont {A.~M.}\ \bibnamefont {Lawrence}}, \bibinfo {author}
  {\bibfnamefont {S.}~\bibnamefont {Weidt}}, \bibinfo {author} {\bibfnamefont
  {F.}~\bibnamefont {Mintert}},\ and\ \bibinfo {author} {\bibfnamefont {W.~K.}\
  \bibnamefont {Hensinger}},\ }\href@noop {} {\bibfield  {journal} {\bibinfo
  {journal} {Physical review letters}\ }\textbf {\bibinfo {volume} {121}},\
  \bibinfo {pages} {180501} (\bibinfo {year} {2018})}\BibitemShut {NoStop}%
\bibitem [{\citenamefont {M{\o}lmer}\ and\ \citenamefont
  {S{\o}rensen}(1999)}]{molmer1999multiparticle}%
  \BibitemOpen
  \bibfield  {author} {\bibinfo {author} {\bibfnamefont {K.}~\bibnamefont
  {M{\o}lmer}}\ and\ \bibinfo {author} {\bibfnamefont {A.}~\bibnamefont
  {S{\o}rensen}},\ }\href@noop {} {\bibfield  {journal} {\bibinfo  {journal}
  {Physical Review Letters}\ }\textbf {\bibinfo {volume} {82}},\ \bibinfo
  {pages} {1835} (\bibinfo {year} {1999})}\BibitemShut {NoStop}%
\bibitem [{\citenamefont {S{\o}rensen}\ and\ \citenamefont
  {M{\o}lmer}(1999)}]{sorensen1999quantum}%
  \BibitemOpen
  \bibfield  {author} {\bibinfo {author} {\bibfnamefont {A.}~\bibnamefont
  {S{\o}rensen}}\ and\ \bibinfo {author} {\bibfnamefont {K.}~\bibnamefont
  {M{\o}lmer}},\ }\href@noop {} {\bibfield  {journal} {\bibinfo  {journal}
  {Physical review letters}\ }\textbf {\bibinfo {volume} {82}},\ \bibinfo
  {pages} {1971} (\bibinfo {year} {1999})}\BibitemShut {NoStop}%
\bibitem [{\citenamefont {Hayes}\ \emph {et~al.}(2012)\citenamefont {Hayes},
  \citenamefont {Clark}, \citenamefont {Debnath}, \citenamefont {Hucul},
  \citenamefont {Inlek}, \citenamefont {Lee}, \citenamefont {Quraishi},\ and\
  \citenamefont {Monroe}}]{hayes2012coherent}%
  \BibitemOpen
  \bibfield  {author} {\bibinfo {author} {\bibfnamefont {D.}~\bibnamefont
  {Hayes}}, \bibinfo {author} {\bibfnamefont {S.~M.}\ \bibnamefont {Clark}},
  \bibinfo {author} {\bibfnamefont {S.}~\bibnamefont {Debnath}}, \bibinfo
  {author} {\bibfnamefont {D.}~\bibnamefont {Hucul}}, \bibinfo {author}
  {\bibfnamefont {I.~V.}\ \bibnamefont {Inlek}}, \bibinfo {author}
  {\bibfnamefont {K.~W.}\ \bibnamefont {Lee}}, \bibinfo {author} {\bibfnamefont
  {Q.}~\bibnamefont {Quraishi}},\ and\ \bibinfo {author} {\bibfnamefont
  {C.}~\bibnamefont {Monroe}},\ }\href@noop {} {\bibfield  {journal} {\bibinfo
  {journal} {Physical review letters}\ }\textbf {\bibinfo {volume} {109}},\
  \bibinfo {pages} {020503} (\bibinfo {year} {2012})}\BibitemShut {NoStop}%
\bibitem [{\citenamefont {L{\"o}schnauer}\ \emph {et~al.}(2025)\citenamefont
  {L{\"o}schnauer}, \citenamefont {Mosca~Toba}, \citenamefont {Hughes},
  \citenamefont {King}, \citenamefont {Weber}, \citenamefont {Srinivas},
  \citenamefont {Matt}, \citenamefont {Nourshargh}, \citenamefont {Allcock},
  \citenamefont {Ballance} \emph {et~al.}}]{loschnauer2025scalable}%
  \BibitemOpen
  \bibfield  {author} {\bibinfo {author} {\bibfnamefont {C.}~\bibnamefont
  {L{\"o}schnauer}}, \bibinfo {author} {\bibfnamefont {J.}~\bibnamefont
  {Mosca~Toba}}, \bibinfo {author} {\bibfnamefont {A.}~\bibnamefont {Hughes}},
  \bibinfo {author} {\bibfnamefont {S.}~\bibnamefont {King}}, \bibinfo {author}
  {\bibfnamefont {M.}~\bibnamefont {Weber}}, \bibinfo {author} {\bibfnamefont
  {R.}~\bibnamefont {Srinivas}}, \bibinfo {author} {\bibfnamefont
  {R.}~\bibnamefont {Matt}}, \bibinfo {author} {\bibfnamefont {R.}~\bibnamefont
  {Nourshargh}}, \bibinfo {author} {\bibfnamefont {D.}~\bibnamefont {Allcock}},
  \bibinfo {author} {\bibfnamefont {C.}~\bibnamefont {Ballance}}, \emph
  {et~al.},\ }\href@noop {} {\bibfield  {journal} {\bibinfo  {journal} {PRX
  Quantum}\ }\textbf {\bibinfo {volume} {6}},\ \bibinfo {pages} {040313}
  (\bibinfo {year} {2025})}\BibitemShut {NoStop}%
\bibitem [{\citenamefont {Tyler~Sutherland}\ and\ \citenamefont
  {Foss-Feig}(2024)}]{tyler2024laser}%
  \BibitemOpen
  \bibfield  {author} {\bibinfo {author} {\bibfnamefont {R.}~\bibnamefont
  {Tyler~Sutherland}}\ and\ \bibinfo {author} {\bibfnamefont {M.}~\bibnamefont
  {Foss-Feig}},\ }\href@noop {} {\bibfield  {journal} {\bibinfo  {journal} {New
  Journal of Physics}\ }\textbf {\bibinfo {volume} {26}},\ \bibinfo {pages}
  {013013} (\bibinfo {year} {2024})}\BibitemShut {NoStop}%
\bibitem [{\citenamefont {Hughes}\ \emph {et~al.}(2025)\citenamefont {Hughes},
  \citenamefont {Srinivas}, \citenamefont {L{\"o}schnauer}, \citenamefont
  {Knaack}, \citenamefont {Matt}, \citenamefont {Ballance}, \citenamefont
  {Malinowski}, \citenamefont {Harty},\ and\ \citenamefont
  {Sutherland}}]{hughes2025trapped}%
  \BibitemOpen
  \bibfield  {author} {\bibinfo {author} {\bibfnamefont {A.}~\bibnamefont
  {Hughes}}, \bibinfo {author} {\bibfnamefont {R.}~\bibnamefont {Srinivas}},
  \bibinfo {author} {\bibfnamefont {C.}~\bibnamefont {L{\"o}schnauer}},
  \bibinfo {author} {\bibfnamefont {H.}~\bibnamefont {Knaack}}, \bibinfo
  {author} {\bibfnamefont {R.}~\bibnamefont {Matt}}, \bibinfo {author}
  {\bibfnamefont {C.}~\bibnamefont {Ballance}}, \bibinfo {author}
  {\bibfnamefont {M.}~\bibnamefont {Malinowski}}, \bibinfo {author}
  {\bibfnamefont {T.}~\bibnamefont {Harty}},\ and\ \bibinfo {author}
  {\bibfnamefont {R.}~\bibnamefont {Sutherland}},\ }\href@noop {} {\bibfield
  {journal} {\bibinfo  {journal} {arXiv preprint arXiv:2510.17286}\ } (\bibinfo
  {year} {2025})}\BibitemShut {NoStop}%
\bibitem [{\citenamefont {Gaebler}\ \emph {et~al.}(2016)\citenamefont
  {Gaebler}, \citenamefont {Tan}, \citenamefont {Lin}, \citenamefont {Wan},
  \citenamefont {Bowler}, \citenamefont {Keith}, \citenamefont {Glancy},
  \citenamefont {Coakley}, \citenamefont {Knill}, \citenamefont {Leibfried}
  \emph {et~al.}}]{gaebler2016high}%
  \BibitemOpen
  \bibfield  {author} {\bibinfo {author} {\bibfnamefont {J.~P.}\ \bibnamefont
  {Gaebler}}, \bibinfo {author} {\bibfnamefont {T.~R.}\ \bibnamefont {Tan}},
  \bibinfo {author} {\bibfnamefont {Y.}~\bibnamefont {Lin}}, \bibinfo {author}
  {\bibfnamefont {Y.}~\bibnamefont {Wan}}, \bibinfo {author} {\bibfnamefont
  {R.}~\bibnamefont {Bowler}}, \bibinfo {author} {\bibfnamefont {A.~C.}\
  \bibnamefont {Keith}}, \bibinfo {author} {\bibfnamefont {S.}~\bibnamefont
  {Glancy}}, \bibinfo {author} {\bibfnamefont {K.}~\bibnamefont {Coakley}},
  \bibinfo {author} {\bibfnamefont {E.}~\bibnamefont {Knill}}, \bibinfo
  {author} {\bibfnamefont {D.}~\bibnamefont {Leibfried}}, \emph {et~al.},\
  }\href@noop {} {\bibfield  {journal} {\bibinfo  {journal} {Physical review
  letters}\ }\textbf {\bibinfo {volume} {117}},\ \bibinfo {pages} {060505}
  (\bibinfo {year} {2016})}\BibitemShut {NoStop}%
\bibitem [{\citenamefont {Leung}\ \emph {et~al.}(2018)\citenamefont {Leung},
  \citenamefont {Landsman}, \citenamefont {Figgatt}, \citenamefont {Linke},
  \citenamefont {Monroe},\ and\ \citenamefont {Brown}}]{leung2018robust}%
  \BibitemOpen
  \bibfield  {author} {\bibinfo {author} {\bibfnamefont {P.~H.}\ \bibnamefont
  {Leung}}, \bibinfo {author} {\bibfnamefont {K.~A.}\ \bibnamefont {Landsman}},
  \bibinfo {author} {\bibfnamefont {C.}~\bibnamefont {Figgatt}}, \bibinfo
  {author} {\bibfnamefont {N.~M.}\ \bibnamefont {Linke}}, \bibinfo {author}
  {\bibfnamefont {C.}~\bibnamefont {Monroe}},\ and\ \bibinfo {author}
  {\bibfnamefont {K.~R.}\ \bibnamefont {Brown}},\ }\href@noop {} {\bibfield
  {journal} {\bibinfo  {journal} {Physical review letters}\ }\textbf {\bibinfo
  {volume} {120}},\ \bibinfo {pages} {020501} (\bibinfo {year}
  {2018})}\BibitemShut {NoStop}%
\bibitem [{\citenamefont {Brownnutt}\ \emph {et~al.}(2015)\citenamefont
  {Brownnutt}, \citenamefont {Kumph}, \citenamefont {Rabl},\ and\ \citenamefont
  {Blatt}}]{brownnutt2015ion}%
  \BibitemOpen
  \bibfield  {author} {\bibinfo {author} {\bibfnamefont {M.}~\bibnamefont
  {Brownnutt}}, \bibinfo {author} {\bibfnamefont {M.}~\bibnamefont {Kumph}},
  \bibinfo {author} {\bibfnamefont {P.}~\bibnamefont {Rabl}},\ and\ \bibinfo
  {author} {\bibfnamefont {R.}~\bibnamefont {Blatt}},\ }\href@noop {}
  {\bibfield  {journal} {\bibinfo  {journal} {Reviews of modern Physics}\
  }\textbf {\bibinfo {volume} {87}},\ \bibinfo {pages} {1419} (\bibinfo {year}
  {2015})}\BibitemShut {NoStop}%
\bibitem [{\citenamefont {Bruzewicz}\ \emph {et~al.}(2015)\citenamefont
  {Bruzewicz}, \citenamefont {Sage},\ and\ \citenamefont
  {Chiaverini}}]{bruzewicz2015measurement}%
  \BibitemOpen
  \bibfield  {author} {\bibinfo {author} {\bibfnamefont {C.}~\bibnamefont
  {Bruzewicz}}, \bibinfo {author} {\bibfnamefont {J.}~\bibnamefont {Sage}},\
  and\ \bibinfo {author} {\bibfnamefont {J.}~\bibnamefont {Chiaverini}},\
  }\href@noop {} {\bibfield  {journal} {\bibinfo  {journal} {Physical Review
  A}\ }\textbf {\bibinfo {volume} {91}},\ \bibinfo {pages} {041402} (\bibinfo
  {year} {2015})}\BibitemShut {NoStop}%
\bibitem [{\citenamefont {Schwerdt}\ \emph {et~al.}(2024)\citenamefont
  {Schwerdt}, \citenamefont {Peleg}, \citenamefont {Shapira}, \citenamefont
  {Priel}, \citenamefont {Florshaim}, \citenamefont {Gross}, \citenamefont
  {Zalic}, \citenamefont {Afek}, \citenamefont {Akerman}, \citenamefont
  {Stern}, \citenamefont {Kish},\ and\ \citenamefont
  {Ozeri}}]{PhysRevX.14.041017}%
  \BibitemOpen
  \bibfield  {author} {\bibinfo {author} {\bibfnamefont {D.}~\bibnamefont
  {Schwerdt}}, \bibinfo {author} {\bibfnamefont {L.}~\bibnamefont {Peleg}},
  \bibinfo {author} {\bibfnamefont {Y.}~\bibnamefont {Shapira}}, \bibinfo
  {author} {\bibfnamefont {N.}~\bibnamefont {Priel}}, \bibinfo {author}
  {\bibfnamefont {Y.}~\bibnamefont {Florshaim}}, \bibinfo {author}
  {\bibfnamefont {A.}~\bibnamefont {Gross}}, \bibinfo {author} {\bibfnamefont
  {A.}~\bibnamefont {Zalic}}, \bibinfo {author} {\bibfnamefont
  {G.}~\bibnamefont {Afek}}, \bibinfo {author} {\bibfnamefont {N.}~\bibnamefont
  {Akerman}}, \bibinfo {author} {\bibfnamefont {A.}~\bibnamefont {Stern}},
  \bibinfo {author} {\bibfnamefont {A.~B.}\ \bibnamefont {Kish}},\ and\
  \bibinfo {author} {\bibfnamefont {R.}~\bibnamefont {Ozeri}},\ }\href
  {https://doi.org/10.1103/PhysRevX.14.041017} {\bibfield  {journal} {\bibinfo
  {journal} {Phys. Rev. X}\ }\textbf {\bibinfo {volume} {14}},\ \bibinfo
  {pages} {041017} (\bibinfo {year} {2024})}\BibitemShut {NoStop}%
\bibitem [{\citenamefont {Shapira}\ \emph {et~al.}(2023)\citenamefont
  {Shapira}, \citenamefont {Peleg}, \citenamefont {Schwerdt}, \citenamefont
  {Nemirovsky}, \citenamefont {Akerman}, \citenamefont {Stern}, \citenamefont
  {Kish},\ and\ \citenamefont
  {Ozeri}}]{shapira2023fastdesignscalingmultiqubit}%
  \BibitemOpen
  \bibfield  {author} {\bibinfo {author} {\bibfnamefont {Y.}~\bibnamefont
  {Shapira}}, \bibinfo {author} {\bibfnamefont {L.}~\bibnamefont {Peleg}},
  \bibinfo {author} {\bibfnamefont {D.}~\bibnamefont {Schwerdt}}, \bibinfo
  {author} {\bibfnamefont {J.}~\bibnamefont {Nemirovsky}}, \bibinfo {author}
  {\bibfnamefont {N.}~\bibnamefont {Akerman}}, \bibinfo {author} {\bibfnamefont
  {A.}~\bibnamefont {Stern}}, \bibinfo {author} {\bibfnamefont {A.~B.}\
  \bibnamefont {Kish}},\ and\ \bibinfo {author} {\bibfnamefont
  {R.}~\bibnamefont {Ozeri}},\ }\href {https://arxiv.org/abs/2307.09566}
  {\bibinfo {title} {Fast design and scaling of multi-qubit gates in
  large-scale trapped-ion quantum computers}} (\bibinfo {year} {2023}),\
  \Eprint {https://arxiv.org/abs/2307.09566} {arXiv:2307.09566 [quant-ph]}
  \BibitemShut {NoStop}%
\bibitem [{\citenamefont {Zhang}\ \emph {et~al.}(2025)\citenamefont {Zhang},
  \citenamefont {Tang}, \citenamefont {Liu}, \citenamefont {Yuan},
  \citenamefont {Shen}, \citenamefont {Wu},\ and\ \citenamefont
  {Zhang}}]{Zhang.25}%
  \BibitemOpen
  \bibfield  {author} {\bibinfo {author} {\bibfnamefont {W.}~\bibnamefont
  {Zhang}}, \bibinfo {author} {\bibfnamefont {G.}~\bibnamefont {Tang}},
  \bibinfo {author} {\bibfnamefont {K.}~\bibnamefont {Liu}}, \bibinfo {author}
  {\bibfnamefont {X.}~\bibnamefont {Yuan}}, \bibinfo {author} {\bibfnamefont
  {Y.}~\bibnamefont {Shen}}, \bibinfo {author} {\bibfnamefont {Y.}~\bibnamefont
  {Wu}},\ and\ \bibinfo {author} {\bibfnamefont {X.-M.}\ \bibnamefont
  {Zhang}},\ }\href@noop {} {\bibfield  {journal} {\bibinfo  {journal} {arXiv
  preprint arXiv:2501.02847}\ } (\bibinfo {year} {2025})}\BibitemShut {NoStop}%
\bibitem [{\citenamefont {Cheng}\ \emph {et~al.}(2023)\citenamefont {Cheng},
  \citenamefont {Liu}, \citenamefont {Geng}, \citenamefont {Fang},\ and\
  \citenamefont {Peng}}]{cheng2023robust}%
  \BibitemOpen
  \bibfield  {author} {\bibinfo {author} {\bibfnamefont {L.}~\bibnamefont
  {Cheng}}, \bibinfo {author} {\bibfnamefont {S.-C.}\ \bibnamefont {Liu}},
  \bibinfo {author} {\bibfnamefont {L.}~\bibnamefont {Geng}}, \bibinfo {author}
  {\bibfnamefont {Y.-K.}\ \bibnamefont {Fang}},\ and\ \bibinfo {author}
  {\bibfnamefont {L.-Y.}\ \bibnamefont {Peng}},\ }\href@noop {} {\bibfield
  {journal} {\bibinfo  {journal} {Physical Review A}\ }\textbf {\bibinfo
  {volume} {107}},\ \bibinfo {pages} {042617} (\bibinfo {year}
  {2023})}\BibitemShut {NoStop}%
\bibitem [{\citenamefont {de~Fouquieres}\ \emph {et~al.}(2011)\citenamefont
  {de~Fouquieres}, \citenamefont {Schirmer}, \citenamefont {Glaser},\ and\
  \citenamefont {Kuprov}}]{de.11}%
  \BibitemOpen
  \bibfield  {author} {\bibinfo {author} {\bibfnamefont {P.}~\bibnamefont
  {de~Fouquieres}}, \bibinfo {author} {\bibfnamefont {S.~G.}\ \bibnamefont
  {Schirmer}}, \bibinfo {author} {\bibfnamefont {S.~J.}\ \bibnamefont
  {Glaser}},\ and\ \bibinfo {author} {\bibfnamefont {I.}~\bibnamefont
  {Kuprov}},\ }\href@noop {} {\bibfield  {journal} {\bibinfo  {journal}
  {Journal of Magnetic Resonance}\ }\textbf {\bibinfo {volume} {212}},\
  \bibinfo {pages} {412} (\bibinfo {year} {2011})}\BibitemShut {NoStop}%
\bibitem [{\citenamefont {Zhang}\ \emph {et~al.}(2019)\citenamefont {Zhang},
  \citenamefont {Wei}, \citenamefont {Asad}, \citenamefont {Yang},\ and\
  \citenamefont {Wang}}]{zhang.19}%
  \BibitemOpen
  \bibfield  {author} {\bibinfo {author} {\bibfnamefont {X.-M.}\ \bibnamefont
  {Zhang}}, \bibinfo {author} {\bibfnamefont {Z.}~\bibnamefont {Wei}}, \bibinfo
  {author} {\bibfnamefont {R.}~\bibnamefont {Asad}}, \bibinfo {author}
  {\bibfnamefont {X.-C.}\ \bibnamefont {Yang}},\ and\ \bibinfo {author}
  {\bibfnamefont {X.}~\bibnamefont {Wang}},\ }\href@noop {} {\bibfield
  {journal} {\bibinfo  {journal} {npj Quantum Information}\ }\textbf {\bibinfo
  {volume} {5}},\ \bibinfo {pages} {85} (\bibinfo {year} {2019})}\BibitemShut
  {NoStop}%
\bibitem [{\citenamefont {Orozco-Ruiz}\ \emph {et~al.}(2025)\citenamefont
  {Orozco-Ruiz}, \citenamefont {Rehman},\ and\ \citenamefont
  {Mintert}}]{orozco2025generally}%
  \BibitemOpen
  \bibfield  {author} {\bibinfo {author} {\bibfnamefont {M.}~\bibnamefont
  {Orozco-Ruiz}}, \bibinfo {author} {\bibfnamefont {W.}~\bibnamefont
  {Rehman}},\ and\ \bibinfo {author} {\bibfnamefont {F.}~\bibnamefont
  {Mintert}},\ }\href@noop {} {\bibfield  {journal} {\bibinfo  {journal}
  {Physical Review A}\ }\textbf {\bibinfo {volume} {111}},\ \bibinfo {pages}
  {042404} (\bibinfo {year} {2025})}\BibitemShut {NoStop}%
\bibitem [{\citenamefont {Walther}\ \emph {et~al.}(2012)\citenamefont
  {Walther}, \citenamefont {Poschinger}, \citenamefont {Singer},\ and\
  \citenamefont {Schmidt-Kaler}}]{walther2012precision}%
  \BibitemOpen
  \bibfield  {author} {\bibinfo {author} {\bibfnamefont {A.}~\bibnamefont
  {Walther}}, \bibinfo {author} {\bibfnamefont {U.}~\bibnamefont {Poschinger}},
  \bibinfo {author} {\bibfnamefont {K.}~\bibnamefont {Singer}},\ and\ \bibinfo
  {author} {\bibfnamefont {F.}~\bibnamefont {Schmidt-Kaler}},\ }\href@noop {}
  {\bibfield  {journal} {\bibinfo  {journal} {Applied Physics B}\ }\textbf
  {\bibinfo {volume} {107}},\ \bibinfo {pages} {1061} (\bibinfo {year}
  {2012})}\BibitemShut {NoStop}%
\bibitem [{\citenamefont {Mart{\'\i}nez-Garc{\'\i}a}\ \emph
  {et~al.}(2022)\citenamefont {Mart{\'\i}nez-Garc{\'\i}a}, \citenamefont
  {Gerster}, \citenamefont {Vodola}, \citenamefont {Hrmo}, \citenamefont
  {Monz}, \citenamefont {Schindler},\ and\ \citenamefont
  {M{\"u}ller}}]{martinez2022analytical}%
  \BibitemOpen
  \bibfield  {author} {\bibinfo {author} {\bibfnamefont {F.}~\bibnamefont
  {Mart{\'\i}nez-Garc{\'\i}a}}, \bibinfo {author} {\bibfnamefont
  {L.}~\bibnamefont {Gerster}}, \bibinfo {author} {\bibfnamefont
  {D.}~\bibnamefont {Vodola}}, \bibinfo {author} {\bibfnamefont
  {P.}~\bibnamefont {Hrmo}}, \bibinfo {author} {\bibfnamefont {T.}~\bibnamefont
  {Monz}}, \bibinfo {author} {\bibfnamefont {P.}~\bibnamefont {Schindler}},\
  and\ \bibinfo {author} {\bibfnamefont {M.}~\bibnamefont {M{\"u}ller}},\
  }\href@noop {} {\bibfield  {journal} {\bibinfo  {journal} {Physical Review
  A}\ }\textbf {\bibinfo {volume} {105}},\ \bibinfo {pages} {032437} (\bibinfo
  {year} {2022})}\BibitemShut {NoStop}%
\bibitem [{\citenamefont {Ruzic}\ \emph {et~al.}(2022)\citenamefont {Ruzic},
  \citenamefont {Chow}, \citenamefont {Burch}, \citenamefont {Lobser},
  \citenamefont {Revelle}, \citenamefont {Wilson}, \citenamefont {Yale},\ and\
  \citenamefont {Clark}}]{ruzic2022frequency}%
  \BibitemOpen
  \bibfield  {author} {\bibinfo {author} {\bibfnamefont {B.~P.}\ \bibnamefont
  {Ruzic}}, \bibinfo {author} {\bibfnamefont {M.~N.}\ \bibnamefont {Chow}},
  \bibinfo {author} {\bibfnamefont {A.~D.}\ \bibnamefont {Burch}}, \bibinfo
  {author} {\bibfnamefont {D.}~\bibnamefont {Lobser}}, \bibinfo {author}
  {\bibfnamefont {M.~C.}\ \bibnamefont {Revelle}}, \bibinfo {author}
  {\bibfnamefont {J.~M.}\ \bibnamefont {Wilson}}, \bibinfo {author}
  {\bibfnamefont {C.~G.}\ \bibnamefont {Yale}},\ and\ \bibinfo {author}
  {\bibfnamefont {S.~M.}\ \bibnamefont {Clark}},\ }\href@noop {} {\bibfield
  {journal} {\bibinfo  {journal} {arXiv preprint arXiv:2210.02372}\ } (\bibinfo
  {year} {2022})}\BibitemShut {NoStop}%
\end{thebibliography}%
% \onecolumngrid

\end{document}